\documentstyle[11pt,fullpage,citesort,psfig,amssymb]{article}

\def\Tr{{\rm Tr}}
\newcommand{\ket}[1]{|#1\rangle}
\newcommand{\bra}[1]{\langle#1|}

\def\ie{{\it i.e.\ }}
\def\m@th{\mathsurround=0pt }
\def\leftrightarrowfill{$\m@th \mathord\leftarrow \mkern-6mu
 \cleaders\hbox{$\mkern-2mu \mathord- \mkern-2mu$}\hfill
 \mkern-6mu \mathord\rightarrow$}
\def\overleftrightarrow#1{\vbox{\ialign{##\crcr
     \leftrightarrowfill\crcr\noalign{\kern-1pt\nointerlineskip}
     $\hfil\displaystyle{#1}\hfil$\crcr}}}

\begin{document}

\begin{titlepage}
\begin{flushright}
UFIFT-HEP-99-15\\
hep-th/9909141
\end{flushright}

\vskip 2.5cm

\begin{center}
\begin{Large}
{\bf QCD Fishnets Revisited}
\end{Large}

\vskip 2.cm

{\large Klaus Bering\footnote{E-mail  address: {\tt bering@phys.ufl.edu}}, 
Joel S. Rozowsky\footnote{E-mail  address: {\tt rozowsky@phys.ufl.edu}} 
and Charles B. Thorn\footnote{E-mail  address: {\tt thorn@phys.ufl.edu}}}

\vskip 0.5cm

{\it Institute for Fundamental Theory\\
Department of Physics, University of Florida,
Gainesville, FL 32611}

(\today)

\vskip 1.0cm
\end{center}

\begin{abstract}
We look back at early efforts to approximate 
the large $N_c$ Feynman diagrams of QCD as very large fishnet
diagrams. We consider more
carefully the uniqueness of rules for discretizing
$P^+$ and $ix^+$
which fix the fishnet model in the strong 't Hooft
coupling limit, and we offer some refinements
that allow more of the crucial QCD interactions
to be retained in the fishnet approximation.
This new discretization has a better chance to
lead to a physically sensible ``bare QCD string'' model.
Not surprisingly the resulting fishnet diagrams
are both richer in structure and harder to 
evaluate than those considered in older work.
As warm-ups we analyze arbitrarily large fishnets of a
paradigm scalar cubic theory and very small
fishnets of QCD.    
\end{abstract}

\vskip 1.0cm
\begin{flushright} 
Copyright 2000 by The American Physical Society
\end{flushright}

\vfill
\end{titlepage}

\renewcommand{\theequation}{\arabic{equation}}
\section{Introduction}
\label{sec1}
With all the recent effort devoted to the search for a
solution of large $N_c$ QCD  \cite{thooftlargen} as a classical string
theory \cite{gubserkp}, it is appropriate to reassess earlier efforts
to accomplish this goal. In this article we wish
to refine and extend the formulation and
calculational methods developed
in the effort of the late seventies to systematize
a fishnet \cite{nielsenfishnet} approximation to large $N_c$ QCD
\cite{thornfishnet,browergt}.
 
The larger goal here is to set up a discrete model of
infinite $N_c$ QCD which, when analyzed in a weak 
coupling expansion ($N_cg^2\ll 1$), reproduces perturbative QCD and,
when analyzed in a strong coupling limit ($N_cg^2\to\infty$), describes
what we choose to call a ``bare QCD string''. Since QCD is supposed to
confine at all values of the 't Hooft coupling, the
infinite $N_c$ glueball should actually be 
noninteracting over the whole range of couplings.
However, its composite internal structure is generally
expected to be quite complicated, and it is only in
the strong 't Hooft coupling limit that the internal
structure of an infinite $N_c$ glueball can be as 
simple as that of the ``fundamental string''
of string theory. We regard it as an open question whether
the bare QCD string can be identified with one
of the known fundamental strings or is an entirely
novel object. We hope that our efforts will eventually
settle this issue. We tentatively identify the
bare QCD string with the object whose propagation is described
by the so-called fishnet diagrams.

As shown in  \cite{thornfishnet} the fishnet
diagrams by no means exhaust the planar
diagrams of 't Hooft's $N_c\to\infty$ limit. 
Fishnets are certainly planar, but they are
also very large in both directions: there are many
lines and many interaction vertices. It is natural to
try to associate such diagrams with strong 't Hooft
coupling $N_cg^2\to\infty$, but as with all strong
coupling expansions one must first define a cutoff
theory which controls the size of the kinetic energy
of the system. In  \cite{thornfishnet} the choice
was to evaluate all graphs on a light front
and simultaneously discretize $P^+=(P^0+P^3)/\sqrt{2}$ and 
$x^+=(x^0+x^3)/\sqrt{2}$.
In this model the large 't Hooft coupling limit
singled out large fishnet diagrams whose continuum
limit was a seamless world sheet. As usual
with strong coupling limits this conclusion is
highly sensitive to the cutoff model. 

What one hopes when resorting to strong coupling methods is that
although the limit strongly distorts the quantitative
details of the continuum theory, the qualitative
physics is shared between the continuum and lattice models
for all couplings.
In standard lattice gauge theories this hope is usually
expressed as requiring that the lattice model exhibit
no phase transition as the coupling constant is
varied from strong to weak coupling. Probably the most
familiar case in which there is a phase transition
is ``compact QED'' whose strong coupling limit shows
confinement, but whose continuum limit is a theory
of free photons.

Although the existence of a phase transition at finite
coupling is usually extremely difficult to detect, it
is the case that in some situations our lattice fishnets
can be seen to be completely irrelevant to the physics
of large but finite coupling. In  \cite{thornfishnet}
this possibility was noted in the context of scalar
$\lambda\phi^4$ theory. The qualitative physics of the
seamless fishnet diagrams is that the quanta of the
field theory are bound into a linear polymeric chain.
However, one can examine at next order in the strong
coupling expansion the nature of the interaction that should
be responsible for this binding. For $\lambda>0$ this 
interaction is {\it repulsive}, and the seamless world
sheet given by the strong coupling limit is a purely
formal artifact. In contrast, for $\lambda<0$ the
interaction is attractive, and it is qualitatively
correct to imagine that the nearest neighbor quanta
form very tight bonds in the strong coupling limit.

A serious shortcoming of the QCD fishnet model attempted
in \cite{browergt} is that the basic gluon-gluon 
quartic interaction
retained in the strong coupling limit favored the alignment
of the gluon spins. This defect was not apparent, however,
because the leading fishnet structure was explicitly an
even function of this interaction and in fact described
an anti-ferromagnetic spin arrangement. The inherent instability
of the system would only be seen at non-leading order.
We aim to improve this situation in the present work
by proposing a discretized model whose fishnet approximation
retains both the ``contact'' interactions, with their
ferromagnetic tendency, and the one gluon exchange
interactions with their anti-ferromagnetic tendency.

The rest of the article is organized as follows. In Section
\ref{sec2} we give a self contained review of the 
Feynman rules in light-cone gauge as well as the
discretization rules largely following \cite{thornfishnet,browergt}.
However, we treat the longitudinal modes differently.
We represent the ``induced quartic interaction'' of 
light-cone gauge by the exchange of a short-lived fictitious spin 0
quantum: its propagation is limited to a small number of 
discrete time steps. This idea motivated another departure
from \cite{browergt}. Namely, we also choose to represent the
basic quartic gluon interaction by the exchange of another
short-lived fictitious spin 0 quantum. In this way
{\it all} vertices of the discretized Feynman rules are
cubic, and are accordingly all treated on the same footing
in the strong coupling limit. In Section \ref{sec3} we
show our discretization in action by computing the
gluon self energy at one loop order. We see how
the ambiguities inherent in  spreading out the
quartic vertices begin to be resolved by requiring
Lorentz invariance. The propagators of the fictitious
scalars are multiplied by $f_k$, $h_k$, where $k$
is the number of time steps, and  $\sum_k f_k=\sum_k h_k=1$.
Lorentz invariance of the self-energy constrains
the moments $\sum_k f_k/k$ and $\sum_k h_k/k$.
In Section \ref{sec4} we describe the fishnet approximation.
As a warm-up, we give a complete analysis of the
leading fishnet diagrams of a paradigm matrix
scalar field theory $g\Tr\phi^3$. Then we describe the
more complicated situation of QCD. We do not attempt
to analyze the arbitrary QCD fishnets here. Instead
in Section \ref{sec5} we study the sum of
planar diagrams for small values of $M$, the
number of units of $P^+$ carried by the evolved
system. Finally in Section \ref{sec6} we collect
some concluding remarks and sketch future
directions for this program of research.

\section{Light-cone Feynman Rules and Discretization}
\label{sec2}
\subsection{Propagators}
The gluon propagator in light-cone gauge $A_-=0$ is given in
momentum space by
\begin{eqnarray}
{\tilde D}^{\mu\nu}(p)
=-i{\eta^{\mu\nu}-\eta^{\mu+}p^\nu/p^+ - \eta^{+\nu}p^\mu/p^+\over
p^2-i\epsilon}.
\end{eqnarray}
The signature of our metric tensor $\eta_{\mu\nu}$ is taken
to be $(-,+,+,+)$.
In this paper we shall make extensive use of the $x^+$ representation
\begin{eqnarray}
D^{\mu\nu}({\bf p}, p^+, x^+)
\equiv\int{dp^-\over2\pi}{\tilde D}^{\mu\nu}(p)e^{-ix^+p^-}.
\end{eqnarray}
Evaluating the $p^-$ integral leads to the following expressions for
the individual components of $D^{\mu\nu}$:
\begin{eqnarray}
D^{kl}({\bf p}, p^+, x^+)&=&\theta(x^+){\eta^{kl}\over 2p^+}
e^{-ix^+{\bf p}^2/2p^+}\, \to\, \theta(\tau){\eta^{kl}\over 2p^+}
e^{-\tau{\bf p}^2/2p^+}
\nonumber\\
D^{k-}({\bf p}, p^+, x^+)&=&\theta(x^+){p^k
\over 2p^{+2}}e^{-ix^+{\bf p}^2/2p^+}\,\to\,\theta(\tau){p^k
\over 2p^{+2}}e^{-\tau{\bf p}^2/2p^+}
\nonumber\\
D^{--}({\bf p}, p^+, x^+)&=&i{\partial\over\partial x^+}
\theta(x^+){1\over p^{+2}}e^{-ix^+{\bf p}^2/2p^+}\,\to\,
-{\partial\over\partial \tau}
\theta(\tau){1\over p^{+2}}e^{-\tau{\bf p}^2/2p^+}\nonumber\\
&=&
\left[\theta(x^+){{\bf p}^2\over2 p^{+3}}
+i\delta(x^+){1\over p^{+2}}\right]e^{-ix^+{\bf p}^2/2p^+}\nonumber\\
&&\qquad\to\,\left[\theta(\tau){{\bf p}^2\over2 p^{+3}}
-\delta(\tau){1\over p^{+2}}\right]e^{-\tau{\bf p}^2/2p^+},
\end{eqnarray}
where the arrows indicate the imaginary time versions ($\tau=ix^+$).
In this paper latin indices will always refer to the transverse
components.
We shall also find it convenient to use a complex basis for transverse indices,
defining $V^\wedge\equiv(V^1+iV^2)/\sqrt2$ and 
$V^\vee\equiv(V^1-iV^2)/\sqrt2$. In this basis the metric
has values $\eta_{\vee\vee}=\eta_{\wedge\wedge}=0$ and
$\eta_{\vee\wedge}=\eta_{\wedge\vee}=1$.

Here we are assuming that $A^-=-A_+$ has not been eliminated from
the formalism. Since Gauss' law relates $A_+$ to the transverse
components through a constraint not involving time derivatives,
it is possible to explicitly integrate $A_+$ out (see for example 
\cite{koguts}) leaving the
transverse components as the only independent variables. In that
case the Feynman rules would only employ the transverse 
propagator $D^{kl}$. Graphically, one achieves the same result
by showing that the contributions of $D^{k-}$ and $D^{--}$
lead to modified cubic vertices and a new induced quartic vertex
which arises from the $\delta(x^+)$ term in $D^{--}$. 

\subsection{Vertices}

\begin{figure}[ht]
\begin{center}
\begin{tabular}{|c|c||c|c|}
\hline
&&&\\[-.3cm]
$
\begin{array}[c]{c}
\psfig{file=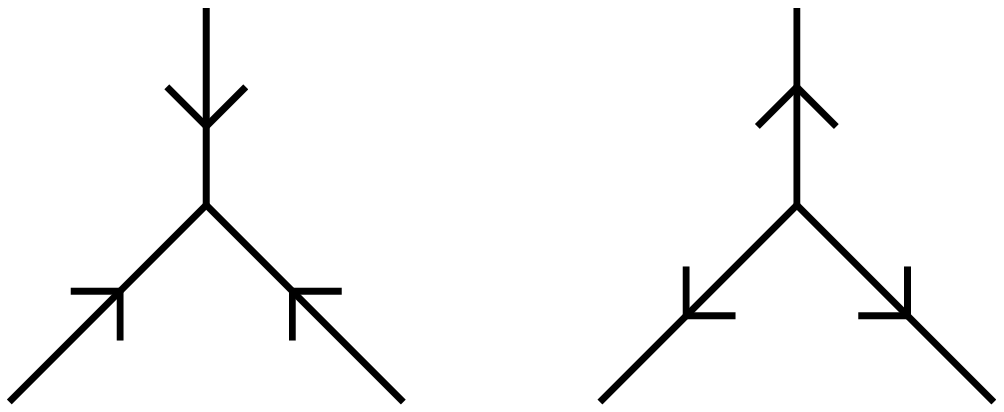,height=0.5in}
\end{array}
$
& 
$\Gamma_3^{\wedge\wedge\wedge} = \Gamma_3^{\vee\vee\vee}=0$
&
$
\begin{array}[c]{c}
\psfig{file=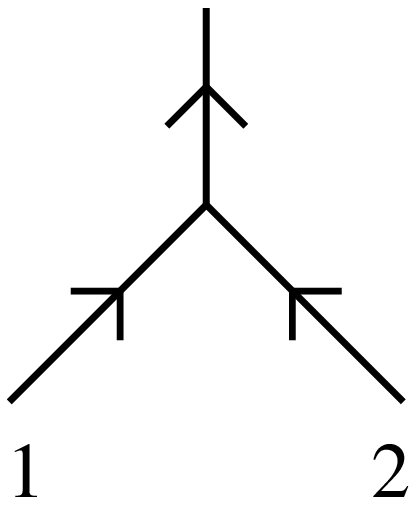,height=0.6in}
\end{array}
$
& 
$ \Gamma_3^{\wedge\wedge\vee} = -g(Q_2-Q_1)^\wedge$ \\
\hline
&&&\\[-.3cm]
$
\begin{array}[c]{c}
\psfig{file=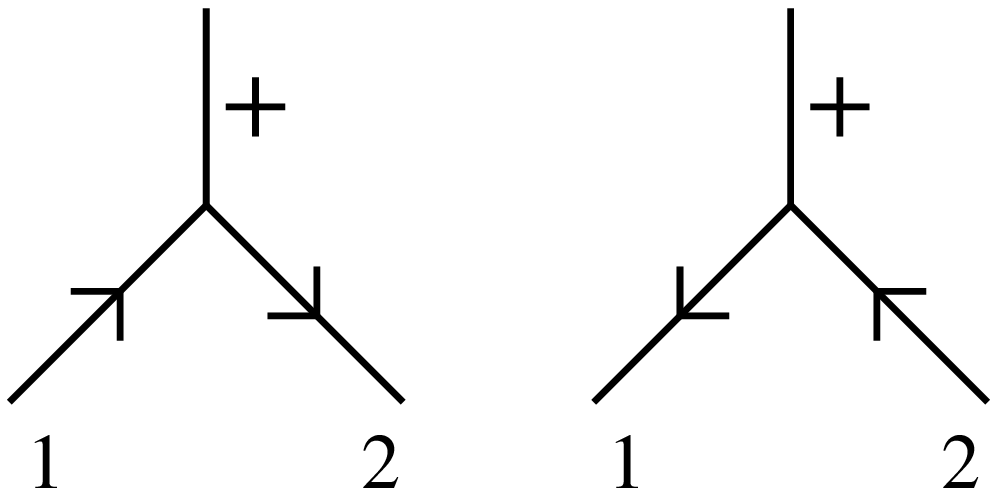,height=0.6in}
\end{array}
$
& 
$ \Gamma_3^{\wedge\vee +} = \Gamma_3^{\vee\wedge +} = g(Q_2-Q_1)^+$ 
&
$
\begin{array}[c]{c}
\psfig{file=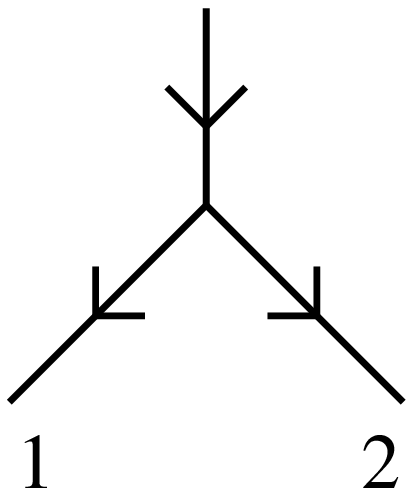,height=0.6in}
\end{array}
$
& 
$\Gamma_3^{\vee\vee\wedge} = -g(Q_2-Q_1)^\vee$ \\
\hline
&&&\\[-.3cm]
$
\begin{array}[c]{c}
\psfig{file=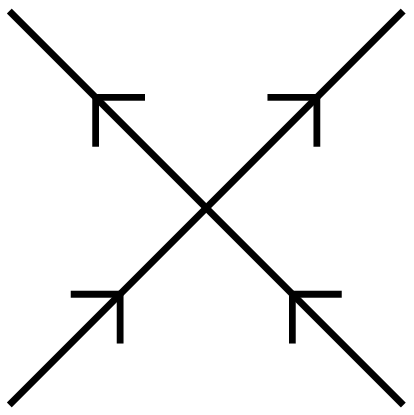,height=0.5in}
\end{array}
$
&
$\Gamma_4^{\wedge\wedge\vee\vee} = +g^2$ 
&
$
\begin{array}[c]{c}
\psfig{file=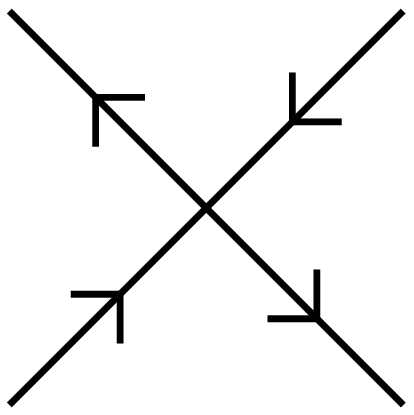,height=0.5in}
\end{array}
$
& 
$\Gamma_4^{\wedge\vee\wedge\vee}  = -2g^2$\\
\hline
\end{tabular}
\end{center}
\caption[]{Cubic and quartic gauge vertices for imaginary time.}
\label{GaugeVertices}
\end{figure}

We shall present the primitive cubic and quartic vertices as 't Hooft did
in his presentation of the $1/N_c$ expansion  \cite{thooftlargen}.
Since the vertex assignments lack permutation symmetry, it is understood that
all permutations of them must be included. The double line
notation makes clear what powers of $N_c$ must be included
with each topology. However, since we shall be dealing exclusively
with the planar diagrams of the $N_c\to\infty$ limit, we shall
dispense with this refinement in order to reduce clutter. To
correctly use these rules at finite $N_c$, the double line 
notation should be restored. With this understanding, the
primitive cubic vertices are given by
\begin{eqnarray}
\Gamma_3^{\vee\vee\vee}&=&\Gamma_3^{\wedge\wedge\wedge}=0\nonumber\\
\Gamma_3^{\wedge\wedge\vee}&=&-ig(Q_2-Q_1)^\wedge\to-g(Q_2-Q_1)^\wedge
\nonumber\\
\Gamma_3^{\vee\vee\wedge}&=&-ig(Q_2-Q_1)^\vee\to-g(Q_2-Q_1)^\vee\nonumber\\
\Gamma_3^{\vee\wedge+}&=&\Gamma_3^{\wedge\vee+}=
+ig(Q_2-Q_1)^+\to+g(Q_2-Q_1)^+,
\end{eqnarray}
and the primitive quartic vertices by
\begin{eqnarray}
\Gamma_4^{\wedge\wedge\vee\vee}&=&+ig^2\to+g^2\nonumber\\
\Gamma_4^{\wedge\vee\wedge\vee}&=&-2ig^2\to-2g^2,
\end{eqnarray}
where the arrows indicate the appropriate vertices to use
with imaginary time.
In light-cone gauge, only transverse gluons participate in the quartic 
vertices. 
Our convention will be that all momenta flow into the vertex.  Also,
the index $\vee$ ($\wedge$) will be represented graphically by
attaching an outgoing (incoming) arrow to a line (see
Fig.~\ref{GaugeVertices}). The ordering of indices will be
counterclockwise around the vertex.
As usual, each vertex is associated with an integration over $x^+$
and conserves the transverse and $+$ components
of momentum. Then each unconstrained momentum is integrated
with measure $d^2pdp^+/(2\pi)^3$.

\subsection{Discretization}
To give a nonperturbative model
for the summation over planar diagrams, it was
proposed in Ref. \cite{thornfishnet} 
to simultaneously discretize $p^+=lm$ and
imaginary time $\tau\equiv ix^+=ka$, with $k,l$ running over all
positive integers. The use of imaginary time converts all
oscillating exponentials to damped ones, and removes all
$i$'s from the Feynman rules. Thus the $i$ occurring in each
vertex is combined with the $dx^+$ to form $d\tau$. (Because
of time translational invariance, the time integral for
one vertex in each connected diagram should be omitted, 
leaving one factor of $i$
unabsorbed. Conventionally, we shall omit this last $i$
in the quantities we calculate.) 

It can be
easily seen that the lattice constants $a,m$ only enter
the sum of graphs in the ratio $T_0\equiv m/a$. First notice
that only this ratio appears in the exponents. Since each
propagator is nominally integrated over its $p^+$,
$\int dp^+\to m\sum$ provides a factor of $m$ to cancel
one from a $1/p^+$ prefactor in each propagator. Further
$\int d\tau\to a\sum$ at each vertex which also has a nominal
$p^+$ conserving delta function that supplies a $1/m$, so
each vertex supplies a factor $1/T_0$. Finally, every $-$
index of a propagator will be matched with a $+$ index of
a vertex, which will involve a factor of $p^+$ and hence
supply a factor of $m$ to cancel the extra factor of $1/m$
in $D^{k-}$ and to convert the extra factor of $1/ma$ in $D^{--}$
to $T_0$.

The discretization of $D^{--}$ involves some ambiguity in
the interpretation of the term involving $\delta(x^+)$. With
$x^+$ continuous, this term collapses the two cubic vertices
it connects into an instantaneous quartic interaction
local in ${\bf x}$ but $p^+$ dependent and hence nonlocal
in $x^-$. Indeed this
is precisely the well-known quartic vertex induced by elimination  of
$A_+$ in the Hamiltonian formulation of light-cone gauge.
In this approach the remaining part of $D^{--}$ 
combines nicely with the contributions of
$D^{k-}$ to yield a modified cubic vertex for transverse
gluons only:
\begin{eqnarray}
{\hat\Gamma}_3^{\wedge\wedge\vee}=-2g
\left({Q^+_1+Q^+_2\over Q^+_1Q^+_2}\right)
(Q^+_1Q^\wedge_2-Q^+_2Q^\wedge_1)
\nonumber\\
{\hat\Gamma}_3^{\vee\vee\wedge}=-2g
\left({Q^+_1+Q^+_2\over Q^+_1Q^+_2}\right)(Q^+_1Q^\vee_2-Q^+_2Q^\vee_1),
\label{modcubic}
\end{eqnarray}
which are the vertices appropriate to imaginary time.
With longitudinal gauge fields completely 
eliminated in this way, discretization
could then proceed as usual by discretizing the $x^+$ and $p^+$
parameters of the transverse gluon propagator. In addition,
one has to exclude, in some {\it ad hoc} manner, 
the $p^+=0$ exchange part of the induced
quartic vertex which is infinite as it stands.
Moreover, the set of ``tadpole diagrams'' necessarily
excluded in our discretization (see Sec.~\ref{tadpolesec} below)
is enlarged by the induced quartic interaction and, since the
new quartic interactions depend non-trivially on $p^+$,
the $p^+$ dependence of the necessary counter-terms will 
be more complicated. 
Nevertheless, an attractive feature of such a treatment is that the
modified cubic vertex is manifestly invariant under the
light-cone Galilei group: ${\bf Q}\to {\bf Q}+Q^+{\bf V}$.

We shall follow a different path, more in the spirit of the sum over
histories. The idea is to exploit our discretization of $x^+$ to
give a more flexible interpretation of $\delta(x^+)$, which
retains a Gaussian damping factor and maintains Galilei invariance throughout.
This can be done by the replacement: 
\begin{eqnarray}
D^{--}({\bf Q}, Q^+=Mm, x^+=-ika)&\to&
{{\bf Q}^2\over 2M^{3}}e^{-k{\bf Q}^2/2MT_0}
-f_k{T_0\over M^{2}}e^{-k{\bf Q}^2/2MT_0}\label{d--3} \nonumber \\
\sum_{k>0}f_k &=&1.
\label{effnorm}
\end{eqnarray}
The last term on the r.h.s.\ of  Eq (\ref{d--3}) 
is a satisfactory discretization of the
delta function provided the $f_k$'s fall off sufficiently
rapidly with $k$. The exclusion of $f_0$ ensures damping
of transverse momentum integrals. 
In this approach, instead of a new induced
quartic vertex we have introduced a short lived
scalar, whose exchange simulates that vertex in a way
that maintains Galilei
invariance. Further, by leaving the choice of the $f_k$'s open
we might be able to tune their values to cancel unwanted
symmetry violations induced by ultraviolet divergences in the continuum
limit. The first term on the r.h.s.\ of Eq (\ref{d--3}) is exactly what is
needed to complete the modified cubic vertex (\ref{modcubic})
when combining all of the contributions of the longitudinal
gluons. 

Perhaps a more intuitive discretization would be to 
replace the derivative in $D^{--}$ by a discrete difference:
\begin{eqnarray}
D^{--}({\bf Q}, Mm, -ika)&\to&-
{T_0\over M^{2}}\cases{e^{-k{\bf Q}^2/2MT_0}-e^{-(k-1){\bf Q}^2/2MT_0}
& $k>1$\cr
e^{-{\bf Q}^2/2MT_0}& $k=1$},
\label{d--2}
\end{eqnarray}
where the special treatment of the case $k=1$ simulates the $\delta(x^+)$
contribution we know must be there in the continuum. Unfortunately, this
definition violates Galilei invariance because of the term
that propagates only $k-1$ steps in time: Newtonian mass conservation
is temporarily violated. This causes considerable
complications in calculations, but nonetheless
displays interesting features. We will not pursue this
option in the main text, but in an appendix
we shall see that with a suitable counter-term Galilei invariance
of the one loop self energy can be restored in the continuum limit. 

\subsection{Tadpoles}
\label{tadpolesec}
The exclusion of the propagators with zero $\tau$
and zero $p^+$ renders every Feynman integral finite,
making our discretization an effective regulator of
divergences.\footnote{The discrete light-cone quantization (DLCQ)
industry which burgeoned in the mid eighties (for a review 
see~\cite{brodskyppreport}) only exploited $P^+$ discretization
leaving ultraviolet divergences unregulated. Discretization of $ix^+$
has the effect of introducing factors of $e^{-a{\bf p}^2/2P^+}$ (which 
is a popular way to regulate UV divergences) into loop integrals.}
But it also means that certain ``tadpole'' Feynman
diagrams, which involve one or more propagators 
originating and terminating at the same vertex are
excluded. In a theory with at most quartic vertices,
these diagrams are limited to self-energy parts, which
generally require counter-terms to enforce Lorentz
invariance in the continuum limit. 
Thus errors induced by excluding these
diagrams could simply be absorbed in the ultimate value
of the counter-terms.

\begin{figure}[ht]
\centerline{\psfig{file=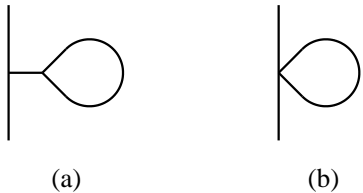,height=1.0in}}
\caption[]{Tadpoles Feynman diagrams coming from cubic and quartic vertices.}
\label{Tadpole1}
\end{figure}
Tadpoles arising from the cubic interaction (see
Fig.~\ref{Tadpole1}(a)) represent the vacuum expectation value of the
color current density.  The transverse components of these would
vanish anyway because they are linear in transverse momentum but we
can make sure all these tadpoles vanish by simply normal ordering the
current density.  However tadpoles arising from the quartic
interaction, see Fig.\ref{Tadpole1}(b), would give a divergent
non-vanishing result if the zero time propagator were inserted. As
mentioned above one possibility is to absorb them in the self-energy
counter-term.

\begin{figure}[ht]
\centerline{\psfig{file=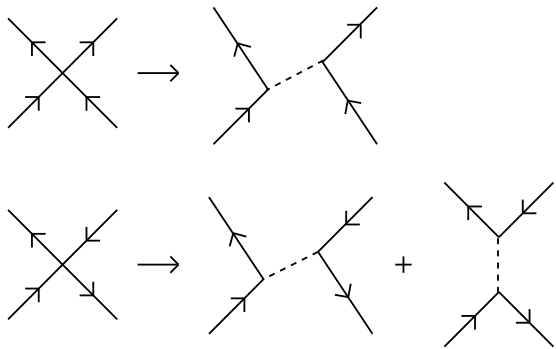,height=1.8in}}
\caption[]{Quartic vertices get replaced by two cubic vertices and a
fictitious scalar field.}
\label{4VertexTo3Vertex}
\end{figure}
\begin{figure}[ht]
\begin{center}
\begin{tabular}{|c|c||c|c|}
\hline
&&&\\[-.3cm]
$
\begin{array}[c]{c}
\psfig{file=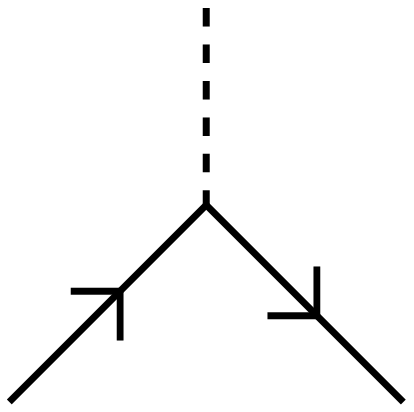,height=0.6in}
\end{array}
$
& 
\hskip .5cm $\Gamma_m^{\wedge\vee} = +{g\over T_0}$ \hskip .5cm 
&
$
\begin{array}[c]{c}
\psfig{file=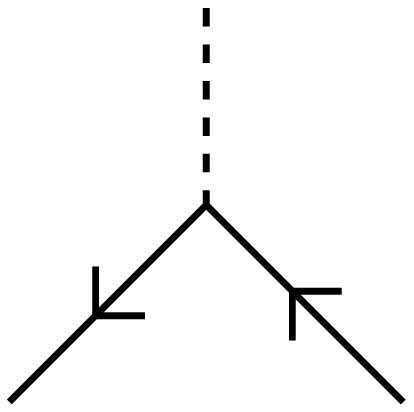,height=0.6in}
\end{array}
$
& 
\hskip .5cm $ \Gamma_m^{\vee\wedge} = -{g\over T_0}$ \hskip .5cm \\
\hline
\end{tabular}
\end{center}
\caption[]{The couplings of the transverse components of the gauge
field to the fictitous scalar field. The subscript $m$ indicates that
this is a {\it magnetic} ghost vertex. These replace the two quartic 
vertices at the bottom of Fig.~\ref{GaugeVertices}.}
\label{FictitousScalarVertex1}
\end{figure}
\begin{figure}[ht]
\centerline{\psfig{file=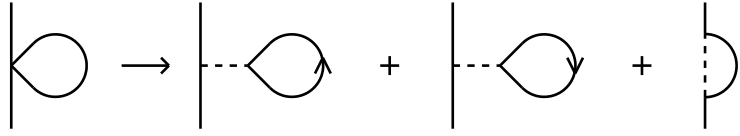,height=0.7in}}
\caption[]{Three tadpole diagrams resulting from the spreading out of
the quartic gauge vertex.}
\label{Tadpole2}
\end{figure}

Another possibility is to note that, from the point of view of the
continuum, one can just as well spread the quartic interaction over
several time steps, in which case a candidate for the tadpole diagrams
would emerge.  We have already exploited this idea in our
discretization of $D^{--}$, see Eq (\ref{d--3}).  A natural way to do
this is to imagine that the quartic interaction is actually the
concatenation of two cubic interactions mediated by a fictitious
scalar field which is only allowed to propagate a few time
steps\footnote{Note that in higher dimensions, the fictitious field
would be a transverse two-form instead of a scalar. Such an additional
degree of freedom is presaged by the first order formulation of gauge
theory in which $F_{\mu\nu}$ is treated as independent of $A_\mu$ and
the Lagrangian density is $-\Tr F^2/4+i\Tr F^{\mu\nu}(\partial_\mu
A_\nu-igA_\mu A_\nu)$.  Going to light-cone gauge in this formalism
leaves, in addition to $A^k$, the (nondynamical) fields $F_{+-}$ and
$F_{kl}$. Our prescription simply gives these extra fields a
short-lived dynamics.}.  We thus redraw the various quartic vertices
as in Fig.~\ref{4VertexTo3Vertex}.  The fictitious scalars must be
``ghosts'': to reproduce the quartic couplings, either their coupling
to two transverse gluons must be taken imaginary or their propagator
must be negative. We choose the second alternative for which the
vertices are given in Fig.~\ref{FictitousScalarVertex1}.
Note that the quartic vertex which involves adjacent like direction
spins in one channel can be viewed as a spin zero exchange in only one
way, whereas the vertex with unlike adjacent spins in both channels
becomes two exchange diagrams, giving a natural interpretation of the
factor of 2 in the effective quartic vertex. If we now consider the
``tadpoles'' arising from connecting any pair of external lines, we
see that there are 3 diagrams, see Fig.~\ref{Tadpole2}, but the two
with the topology of a cubic tadpole, which cannot be
drawn in our discretized light-cone formalism, cancel. Thus the only remaining
tadpole is the bubble diagram on the far right of Fig.~\ref{Tadpole2},
which poses no problem for our formalism.

The fictitious scalar propagator can be taken to be 
\begin{eqnarray}
D({\bf Q},M,k)&=&-h_k T_0 e^{-k{\bf Q}^2/2MT_0} \nonumber \\
\sum_{k>0} h_k&=&1,
\label{hnorm}
\end{eqnarray}
where the $h_k$'s, like the $f_k$'s, vanish rapidly with $k$.
The normalization condition guarantees that the correct
quartic vertex will be reproduced in the continuum limit.
The exponential factor damps ultraviolet divergences, and
with the form we have specified, maintains Galilei invariance
even for finite lattice constants. Furthermore, by leaving
the choice of the $h_k$'s open, we gain additional
flexibility to cancel unwanted symmetry violations.
The hope is that tuning the $f_k$'s and $h_k$'s
will remove the need for explicit counter-terms.

\section{Gluon Self-energy and Counter-terms at One Loop}
\label{sec3}
In this section we illustrate the way discretization regulates
divergences by computing the one-loop contribution to the gluon
self-energy part, $i\Pi^{\mu\nu}$, defined as the sum of all one particle
irreducible diagrams for the two-point function. For this purpose
it is convenient to pass from $x^+$ representation to energy ($E=p^-$)
representation. With discretized $\tau$, this is accomplished by
defining
\begin{eqnarray}
{\tilde D}^{\mu\nu}({\bf Q}, M, E)
&=&\sum_{k=1}^\infty e^{akE}D^{\mu\nu}({\bf Q}, M, k)\\
{\tilde \Pi}^{\mu\nu}({\bf Q}, M, E)
&=&\sum_{k=1}^\infty e^{akE}\Pi^{\mu\nu}({\bf Q}, M, k).
\end{eqnarray}
The exact gluon propagator is then algebraically related to the
the bare one and ${\tilde \Pi}$. We define $u\equiv e^{-Q^2/2MT_0} 
= e^{aE-{\bf Q}^2/2MT_0} $,
in terms of which the bare propagators have the values: 
\begin{eqnarray}
{\tilde D}_0^{ij}({\bf Q}, M, E)&=&{\eta^{ij}\over2M}{u\over 1-u}
\label{transverseprop} \nonumber \\
{\tilde D}_0^{i-}({\bf Q}, M, E)&=&{Q^i\over 2M^{2}}{u\over 1-u}.
\end{eqnarray}
For $D^{--}$ we have, for the discretization of Eq (\ref{d--3}),
\begin{eqnarray}
{\tilde D}_{0}^{--}({\bf Q}, M, E)&=&{{\bf Q}^2\over2M^3}{u\over 1-u}
-{T_0\over M^2}\sum_{k>0}f_ku^k.
\end{eqnarray}

Because of light-cone gauge, only $\Pi^{ij}, \Pi^{i+}$ and $\Pi^{++}$
are required. By transverse rotational invariance, these quantities can
be decomposed as
\begin{eqnarray}
{\tilde\Pi}^{ij}&=&Q^iQ^j\Pi_1+\eta^{ij}\Pi_2\nonumber\\
{\tilde\Pi}^{i+}&=&MQ^i\Pi^\prime_1\nonumber\\
{\tilde\Pi}^{++}&=&M^2\Pi^{\prime\prime}_1
\label{defpi12}
\end{eqnarray}
We shall find that at the one-loop approximation, $\Pi^{\prime\prime}_1
=\Pi^\prime_1=\Pi_1$ and if that were to hold generally, the exact
propagators would be given by
\begin{eqnarray}
{\tilde D}^{ij}&=&{\eta^{ij}\over2M}{u\over 1-u-u\Pi_2/2M}
\nonumber\\
{\tilde D}^{i-}&=&{Q^i\over 2M^{2}}{u\over 1-u-u\Pi_2/2M}
\nonumber\\
{\tilde D}^{--}&=&{T_0\over M^2}\left[{-\sum_kf_ku^k
\over 1+T_0\Pi_1\sum_k f_ku^k}
+{{\bf Q}^2\over2MT_0}{u\over 1-u-u\Pi_2/2M}\right].
\label{exactprops}
\end{eqnarray}
\subsection{Calculation of $\Pi_1$}
\begin{figure}[ht]
\centerline{\psfig{file=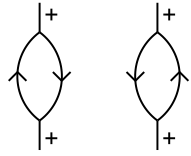,height=0.8in}}
\caption[]{The two bubble diagrams that contribute to $\Pi^{++}$.}
\label{PiPlusPlus}
\end{figure}
After these preliminaries, let us now turn to the computation of
$\Pi_1$ at one loop.  The simplest term is $\Pi^{++}$, which is given
in $x^+$ representation by (note that there are 2 equal diagrams
that contribute (see Fig.~\ref{PiPlusPlus}))
\begin{eqnarray}
\Pi^{++}&=&2{g^2N_c\over T_0^2}
\int{d^2p\over(2\pi)^3}\sum_{l=1}^{M-1}{(M-2l)^2\over
4l(M-l)}e^{-k({\bf p}^2/2l+({\bf Q}-{\bf p})^2/2(M-l))/T_0}
\nonumber\\
&=&{2g^2N_c\over16\pi^2T_0}\sum_{l=1}^{M-1}{(M-2l)^2}{1\over Mk}
e^{-k{\bf Q}^2/2MT_0}={N_cg^2(M-1)(M-2)\over24\pi^2T_0 k}
e^{-k{\bf Q}^2/2MT_0}\\
{\tilde\Pi}^{++}&=&{g^2N_c\over24\pi^2T_0}(M-1)(M-2)(-\ln(1-u)).
\end{eqnarray}

\begin{figure}[ht]
\centerline{\psfig{file=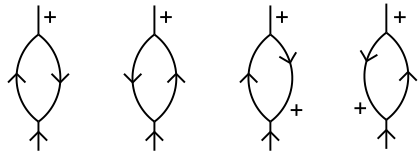,height=0.8in}}
\caption[]{The four bubble diagrams that contribute to $\Pi^{\wedge +}$.}
\label{PiUpPlus}
\end{figure}
The evaluation of $\Pi^{j+}$ is not much harder. Here there are four
diagrams that differ only in the prefactor of transverse momentum (see
Fig.~\ref{PiUpPlus}). Since the prefactor is linear in momentum, we
need only to remember that the Gaussian integral over transverse
momentum involves the shift ${\bf p}\to{\bf p}+l{\bf Q}/M$.  After
this shift the term linear in $\bf p$ integrates to zero, so the net
effect is to set ${\bf p}=l{\bf Q}/M$. Thus the net prefactor is
\begin{eqnarray}
{Q^j\over M}(2l-M)[(1+l/M)+(-2+l/M)+\eta_{+-}(-1-l/M)+
\eta_{+-}(2-l/M)]={2Q^j\over M^2}(2l-M)^2,
\end{eqnarray}
which involves the identical sum over $l$ as $\Pi^{++}$. Thus we end up
with
\begin{eqnarray}
{\tilde\Pi}^{j+}&=&{g^2N_c\over24\pi^2T_0}{Q^j\over M}(M-1)(M-2)(-\ln(1-u)).
\end{eqnarray}

\begin{figure}[ht]
\centerline{\psfig{file=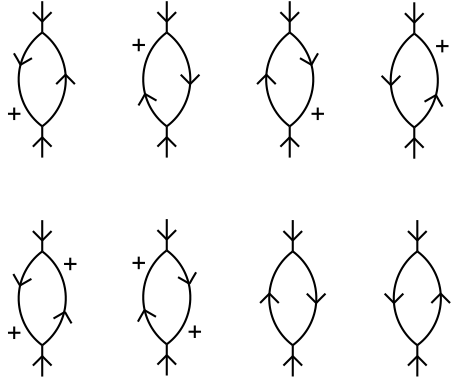,height=2.0in}}
\caption[]{The eight bubble diagrams that 
contribute to $\Pi^{\wedge\wedge}$.}
\label{PiUpUp}
\end{figure}
The coefficient of $Q^{i}Q^j$ in $\Pi^{ij}$ may be singled out by
computing $\Pi^{\wedge\wedge}$. This leads to a similar calculation to
the above because, after the shift in integration variable, $p^\wedge$
and $p^\wedge p^\wedge$ both integrate to zero. Thus again, prefactors
of $p^\wedge$ may simply be set to $lQ^\wedge/M$. Eight diagrams
contribute to this quantity, equal in pairs (see
Fig.~\ref{PiUpUp}). The prefactors of these diagrams combine as
follows:
\begin{eqnarray}
\hskip -0.2cm
{2Q^\wedge Q^\wedge\over M^2}((M+l)^2+(2M-l)^2-(M+l)(2M-l)-(M+l)(2M-l))=
{2Q^\wedge Q^\wedge\over M^2}(2l-M)^2.
\end{eqnarray}
So again we have the same sum over $l$, leading to:
\begin{eqnarray}
\Pi^{\wedge\wedge}={g^2N_c\over24\pi^2T_0}{Q^\wedge Q^\wedge\over M^2}
(M-1)(M-2)(-\ln(1-u)).
\end{eqnarray}
The upshot of the calculations so far is that 
\begin{eqnarray}
\Pi_1=\Pi_1^\prime=\Pi_1^{\prime\prime}={g^2N_c\over24\pi^2T_0}
(1-{3\over M}+{2\over M^2})(-\ln(1-u)).
\end{eqnarray}
The equality of the various $\Pi$'s holds even at finite $m,a$.
This can be understood because the diagrams we have evaluated
show no violation of Galilei invariance. 
Our result for $\Pi_1$ can be compared with the result from the
study of asymptotic freedom in the infinite momentum frame  
\cite{thornfreedom}:
\begin{eqnarray}
\Pi_1(Q^2)={g^2N_c\over24\pi^2}\left(\ln{\Lambda^2\over Q^2}+12\right),
\end{eqnarray}
in which a simple cutoff ${\bf k}^2<\Lambda^2$ was employed. To make the
comparison, note that in the continuum limit, $a\to0,M\to\infty$,
$1-u\approx ({\bf Q}^2-2Q^+E)/2MT_0=Q^2/2MT_0$. Thus for us, the
role of $\Lambda^2$ is played by the quantity $2MT_0=2Q^+/a$. Note 
that if we choose to keep $T_0$ fixed, the ultraviolet cutoff
is removed by simply taking $M\to\infty$.
\subsection{Calculation of $\Pi_2$}
Finally we turn to the real core of the self energy calculation,
the determination of $\Pi_2$, which can be inferred from
$\Pi^{\wedge\vee}=Q^\wedge Q^\vee\Pi_1+\Pi_2=\Pi_2+{\bf Q}^2\Pi_1/2$.
Seventeen diagrams contribute to this quantity, fifteen of
which do not involve $D^{--}$ and are relatively simple to
analyze. 
\begin{figure}[ht]
\centerline{\psfig{file=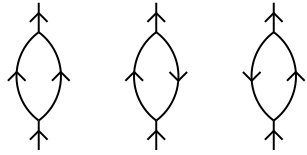,height=0.8in}}
\caption[]{The three bubble diagrams that contribute to
$\Pi^{\wedge\vee}_{\rm A}$ involving only transverse components.}
\label{PiUpDownA}
\end{figure}
First consider the three diagrams that only involve transverse
internal propagators (see Fig.~\ref{PiUpDownA}). The prefactors of transverse
momentum combine as follows:
\begin{eqnarray}
(2p-Q)^\wedge(2p-Q)^\vee+(p+Q)^\wedge(p+Q)^\vee+(2Q-p)^\wedge(2Q-p)^\vee
&=&3({\bf p}^2+{\bf Q}^2-{\bf p}\cdot{\bf Q})\nonumber\\
\to 3({\bf p}^2+(1-l/M+(l/M)^2){\bf Q}^2),&&
\end{eqnarray}
where in the last line we have indicated the result of shifting ${\bf p}$
by $l{\bf Q}/M$ and dropping the term linear in ${\bf p}$.
The transverse integral of the term in ${\bf Q}^2$ is the same
as before, but the term in ${\bf p}^2$ gives an extra factor
$2T_0l(M-l)/kM$. These terms require the sums
\begin{eqnarray}
\sum_{l=1}^{M-1}(M^2-lM+l^2)={M(M-1)(5M-1)\over6},\qquad
\sum_{l=1}^{M-1}l(M-l)={M(M-1)(M+1)\over6}
\end{eqnarray}
respectively. The contribution of these three diagrams, 
$\Pi^{\wedge\vee}_{\rm A}$, is given by
\begin{eqnarray}
\Pi^{\wedge\vee}_{\rm A} = 
{g^2N_c\over16\pi^2}\left[{M^2-1\over k^2M}
+{5M^2-6M+1\over2kM^2T_0}{\bf Q}^2
\right]e^{-k{\bf Q}^2/2MT_0}.
\end{eqnarray}

\begin{figure}[ht]
\centerline{\psfig{file=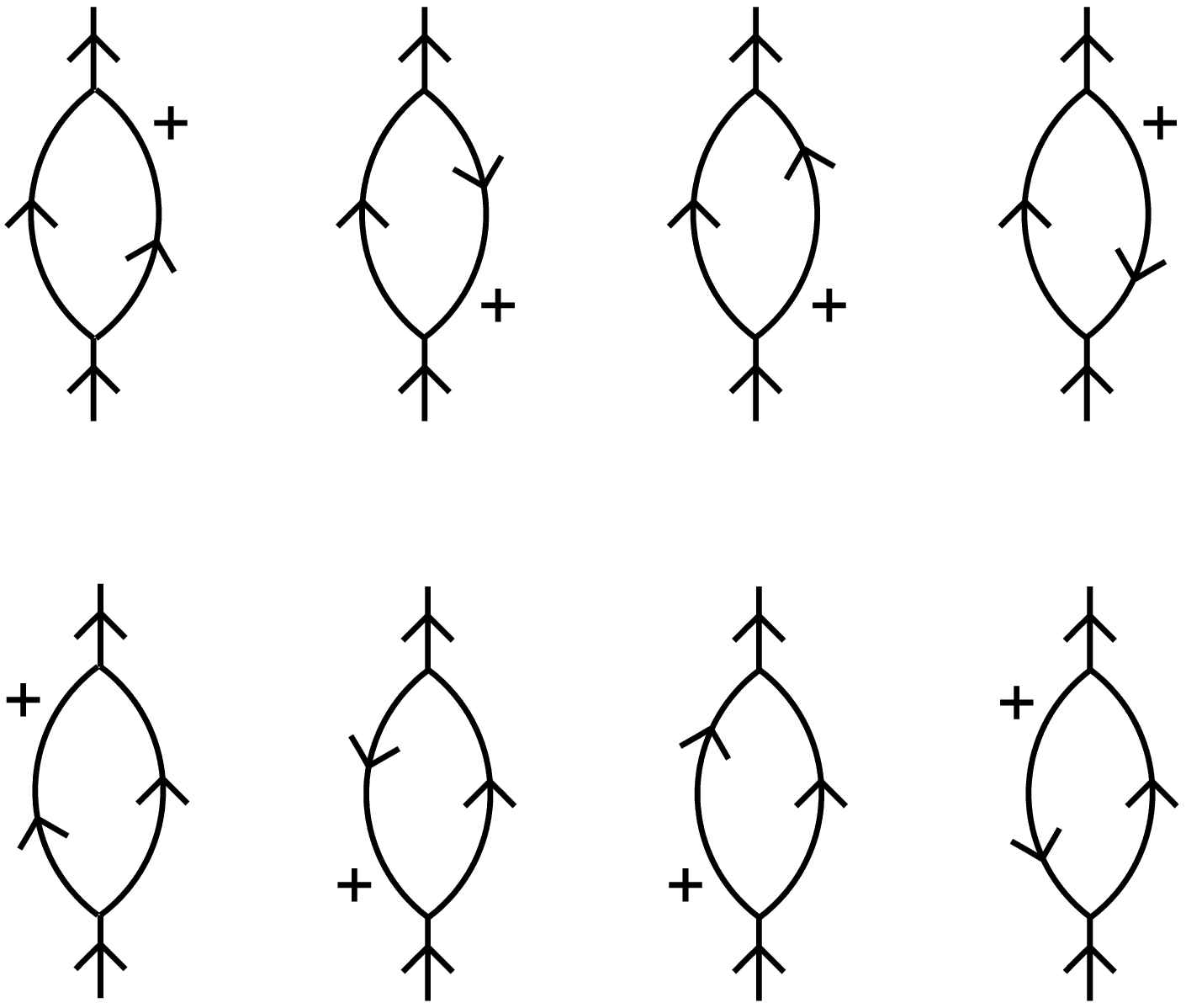,height=2.0in}}
\caption[]{The eight bubble diagrams that contribute to
$\Pi^{\wedge\vee}_{\rm B}$.}
\label{PiUpDownB}
\end{figure}
The next class of diagrams consists of the eight graphs with one
$D^{j+}$ propagator, shown in Fig.~\ref{PiUpDownB}. Remembering the
single factor $\eta_{+-}$, we find that the prefactors combine as follows:
\begin{eqnarray}
&&-(2M-l)\left[{p^\wedge\over l}
(Q-2p+Q+p)^\vee+{p^\vee\over l}(Q-2p+Q+p)^\wedge
\right]\nonumber\\
&&\qquad-(M+l)\left[{(Q-p)^\wedge\over M-l}
(-Q+2p+2Q-p)^\vee+{(Q-p)^\vee\over M-l}
(-Q+2p+2Q-p)^\wedge\right]\nonumber\\
&&\qquad\qquad=-{2M-l\over l}{\bf p}\cdot(2{\bf Q}-{\bf p})-{M+l\over M-l}
({\bf Q}-{\bf p})\cdot({\bf Q}+{\bf p})\nonumber\\
&&\qquad\qquad\to-{5M^2-2lM+2l^2\over M^2}{\bf Q}^2+{2M^2-2lM+2l^2
\over l(M-l)}{\bf p}^2
\end{eqnarray}
where as before the arrow indicates the effect on the prefactors after
the usual shift of integration variables.
Performing the by now familiar integrals and sums leads to the 
following result for the contribution of this class of eight diagrams, 
labeled by $\Pi^{\wedge\vee}_{\rm B}$:
\begin{eqnarray}
\Pi^{\wedge\vee}_{\rm B} = {g^2N_c\over24\pi^2}
\left[{5M^2-6M+1\over k^2M}-{14M^2-15M+1\over2kM^2T_0}{\bf Q}^2
\right]e^{-k{\bf Q}^2/2MT_0}.
\end{eqnarray}
Combining this with the previous three diagrams yields for the eleven
diagrams considered thus far
\begin{eqnarray}
\Pi^{\wedge\vee}_{\rm A} + \Pi^{\wedge\vee}_{\rm B} =
{g^2N_c\over48\pi^2}{13M^2-12M-1\over M}
\left[{1\over k^2}-{{\bf Q}^2\over2kMT_0}
\right]e^{-k{\bf Q}^2/2MT_0}.
\end{eqnarray}

\begin{figure}[ht]
\centerline{\psfig{file=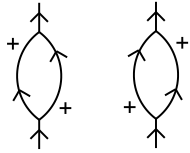,height=0.8in}}
\caption[]{The two bubble diagrams that contribute to
$\Pi^{\wedge\vee}_{\rm C}$.}
\label{PiUpDownC}
\end{figure}
The prefactors in the pair of diagrams (shown in Fig.~\ref{PiUpDownC})
involving two $D^{j+}$ propagators combine to form
\begin{eqnarray}
{(M+l)(l-2M)\over l(M-l)}{\bf p}\cdot({\bf Q}-{\bf p})
\to{(M+l)(2M-l)\over l(M-l)}{\bf p}^2-{(M+l)(2M-l)\over M^2}{\bf Q}^2.
\end{eqnarray}
This time the required sum is just
\begin{eqnarray}
\sum_{l=1}^{M-1}(2M^2+lM-l^2)=M(M-1)(13M+1)/6,
\end{eqnarray}
so that these two diagrams simply double the result of the first 
eleven. So, in summary, the total contributions of the thirteen 
diagrams that do not involve $D^{--}$ to $\Pi^{\wedge\vee}$ and 
${\tilde \Pi}^{\wedge\vee}$ are
\begin{eqnarray}
\Pi^{\wedge\vee}_{\rm A}+\Pi^{\wedge\vee}_{\rm B}+\Pi^{\wedge\vee}_{\rm C}
={g^2N_c\over24\pi^2}
{13M^2-12M-1\over M}\left[{1\over k^2}-{{\bf Q}^2\over2kMT_0}
\right]e^{-k{\bf Q}^2/2MT_0}\\
{\tilde \Pi}^{\wedge\vee}_{\rm A}+{\tilde \Pi}^{\wedge\vee}_{\rm B}+
{\tilde \Pi}^{\wedge\vee}_{\rm C} ={g^2N_c\over24\pi^2}
{13M^2-12M-1\over M}\left[\sum_{k=1}^\infty{u^k\over k^2}
+{{\bf Q}^2\over2MT_0}
\ln(1-u)\right]
\end{eqnarray}

\begin{figure}[ht]
\centerline{\psfig{file=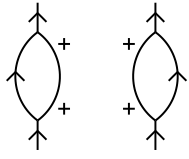,height=0.8in}}
\caption[]{The two bubble diagrams that contribute to
$\Pi^{\wedge\vee}_{\rm D}$ which involve a $D^{--}$ propagator.}
\label{PiUpDownD}
\end{figure}
The two diagrams in Fig.~\ref{PiUpDownD}, have a $D^{--}$
propagator as one of the internal lines and lead to a qualitatively
different evaluation than the first thirteen. 
Using (\ref{d--3}) the Gaussian exponents are the same 
as with the first thirteen diagrams. One finds
\begin{eqnarray}
\Pi^{\wedge\vee}_{\rm D} &=& 
{g^2N_c\over8\pi^2}\sum_{l=1}^{M-1}\left\{{(2M-l)^2\over l}
e^{-k{\bf Q}^2/2MT_0}
\left[{M-l\over k^2M^2}-{f_k\over Mk}
+{l{\bf Q}^2\over2kM^3T_0}\right]
+(l\to M-l)\right\}\qquad\qquad \nonumber\\
&=&{g^2N_c\over4\pi^2}e^{-k{\bf Q}^2/2MT_0}
\bigg[{1-kf_k\over k^2}(4M[\psi(M)+\gamma]-7(M-1)/2)\nonumber\\
& &-{(M-1)(14M-1)\over 6k^2M}+{(M-1)(14M-1){\bf Q}^2\over12kM^2T_0}\bigg], 
\end{eqnarray}
where we have made use of the identity
\begin{eqnarray}
{1\over M}\sum_{l=1}^{M-1}{(2M-l)^2\over l}=4M[\psi(M)+\gamma]
-{7\over2}[M-1],
\label{psisum}
\end{eqnarray}
where $\psi(z)=\Gamma^\prime(z)/\Gamma(z)$ is the digamma function
and $\gamma$ is Euler's constant. At large $M$, we have
\begin{eqnarray}
\psi(M)\sim \ln M-{1\over2M}-\sum_{n\geq1}{B_{2n}\over2nM^{2n}},
\end{eqnarray}
where $B_{2n}$ are the Bernoulli numbers.

\begin{figure}[ht]
\centerline{\psfig{file=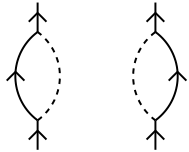,height=0.8in}}
\caption[]{The two bubble diagrams that contribute to
$\Pi^{\wedge\vee}_{\rm E}$ which correspond to the tadpole 
contributions induced by the spread out quartic vertices.}
\label{PiUpDownE}
\end{figure}
With the quartic vertices realized as the exchange of a
magnetic scalar, we have a definite proposal for the
tadpole contribution to the self-energy, namely the two
diagrams of Fig.~\ref{PiUpDownE}, which each give equal contributions. 
Calling the tadpole contribution $\Pi^{\wedge\vee}_E$, we find
\begin{eqnarray}
\Pi^{\wedge\vee}_E&=&-2{g^2N_c\over(2\pi)^3T_0}\int d^2p\sum_{l=1}^{M-1}
{h_k\over2(M-l)}e^{-k[M{\bf p}^2/l(M-l)+{\bf Q}^2/M]/2T_0}\nonumber\\
&=&-{g^2N_c\over4\pi^2}\sum_{l=1}^{M-1}
{lh_k\over Mk}e^{-k{\bf Q}^2/2MT_0}  \nonumber \\
&=&-{g^2N_c\over8\pi^2}(M-1){h_k\over k}e^{-k{\bf Q}^2/2MT_0} 
\label{piE}
\end{eqnarray}
Adding $\Pi^{\wedge\vee}_D+\Pi^{\wedge\vee}_E$ 
to the first thirteen diagrams, gives
(in energy representation)
\begin{eqnarray}
{\tilde \Pi}^{\wedge\vee} 
&=&{\tilde \Pi}^{\wedge\vee}_{\rm A}+{\tilde \Pi}^{\wedge\vee}_{\rm B}+
{\tilde \Pi}^{\wedge\vee}_{\rm C}+{\tilde \Pi}^{\wedge\vee}_{\rm D}
+{\tilde \Pi}^{\wedge\vee}_{\rm E}\nonumber\\
&=&{g^2N_c\over4\pi^2}
\bigg[\sum_{k=1}^\infty{u^k\over k^2}
\left(4M[\psi(M)+\gamma]-{(M-1)(11M-1)\over3M} \right)-
\sum_{k=1}^\infty h_k {u^k\over k}{M-1 \over 2}\nonumber\\
& &-\sum_{k=1}^\infty f_k{u^k\over k}
\left(4M[\psi(M)+\gamma]-{7(M-1)\over2}\right)
+\sum_{k=1}^\infty{u^k\over k}{(M-1)(M-2){\bf Q}^2\over12M^2T_0}\bigg].
\label{pi2notinv}
\end{eqnarray}
To get $\Pi_2$ we must subtract ${\bf Q}^2{\Pi}_1/2$, which exactly
cancels the last term leaving:
\begin{eqnarray}
{\Pi}_2 
&=&{g^2N_c\over4\pi^2}
\bigg[\sum_{k=1}^\infty{u^k\over k^2}
\left(4M[\psi(M)+\gamma]-{(M-1)(11M-1)\over3M}
\right)\nonumber\\
& &
-\sum_{k=1}^\infty h_k{u^k\over k}{M-1\over2}
-\sum_{k=1}^\infty f_k{u^k\over k}
\left(4M[\psi(M)+\gamma]-{7(M-1)\over2}\right)
\bigg].
\label{pi2inv}
\end{eqnarray}
The first term inside the large square brackets is exactly the
contribution one would obtain using the modified cubic vertex
(\ref{modcubic})  in the three diagrams of Fig.~\ref{PiUpDownA}.
The terms with $f_k$ represent the tadpole diagram that
would have been constructed
from the induced quartic vertex had $A_+$ been
eliminated from the formalism. This is as it should be because summing
over all the diagrams involving $D^{-k}$ and $D^{--}$ is
the graphical equivalent of the elimination of the $A_+$ degree
of freedom. The formalism with $A_+$ eliminated gives an extremely
efficient calculation of the three bubble diagrams, as shown in the
next subsection, but no candidate for the tadpoles. 
But by treating $A_+$ graphically
we are led to a useful proposal for the tadpole diagram. 

In order to compare our results to the continuum calculation of
Ref.~\cite{thornfreedom}, we must examine the limits $u\to 1$
and $M\to\infty$. The behavior of the first term of
Eq.~(\ref{pi2inv}) in this limit is transparent once we use the identity
\begin{eqnarray}
\sum_{k=1}^\infty {u^k\over k^2}={\pi^2\over6}-\ln u\ \ln(1-u)
-\sum_{k=1}^\infty {(1-u)^k\over k^2}\approx{\pi^2\over6}+(1-u)(\ln(1-u)-1),
\label{uto1-u}
\end{eqnarray}
for $u\approx1$. The last two terms can be expanded about
$u=1$:
\begin{eqnarray}
\sum_{k=1}^\infty h_k{u^k\over k}&=&\sum_{k=1}^\infty {h_k\over k}
+(u-1)\sum_{k=1}^\infty h_k+{\cal O}((u-1)^2) \nonumber\\
&=&\sum_{k=1}^\infty {h_k\over k}
+(u-1)+{\cal O}((u-1)^2),
\label{hexpansion}
\end{eqnarray}
and a similar expansion for $\sum_k f_k u^k/k$.
The coefficient of the linear term is completely
determined by the normalization conditions on the $f$'s and $h$'s.
Momentum independent terms in $\Pi_2$ imply a (divergent and
noncovariant) tachyonic gluon 
mass squared in perturbation theory. Since the gluon has only
helicity $\pm1$, such a mass is inconsistent with Poincar\'e
invariance of the continuum limit. Clearly these symmetry
violating terms would be cancelled if we could impose the 
constraints
\begin{eqnarray}
\sum_{k=1}^\infty {f_k\over k}&=&{\pi^2\over6} \label{fsum}\\
\sum_{k=1}^\infty{h_k\over k}&=&
-{\pi^2\over18}\left(1-{2\over M}\right),
\label{hsumold}
\end{eqnarray}
but clearly the second of these is impossible with $h_k$ 
independent of $M$ (implicitly assumed in (\ref{hnorm})). 
The best we can do
is to set the r.h.s.\ of Eq.~(\ref{hsumold}) to $-\pi^2/18$, which would
cancel the linear divergence of $\Pi_2$. However, there remains a 
finite gluon mass which is still inconsistent with the Poincar\'e 
invariance of the continuum theory. One option would be to cancel this
with a mass counter-term, whose coefficient would have to be determined 
order by order.

Another approach is to allow $h_k$ to be dependent on $M$. We have the 
freedom to make this replacement as long as we recover the correct 
continuum limit ($M\rightarrow\infty$ and $a\rightarrow 0$) of the 
theory\footnote{We have the same flexibility for the $f$'s
in (\ref{effnorm}) but as we will see this is not necessary at least
at one loop order.}\hskip-.15cm. With this modification, 
Eq.~(\ref{piE}) is replaced by
\begin{equation}\Pi^{\wedge\vee}_E = -{g^2N_c\over4\pi^2}\sum_{l=1}^{M-1}
{lh_k(l)\over Mk}e^{-k{\bf Q}^2/2MT_0},
\label{piEnew}
\end{equation}
where the sum over $l$ is not performed.
Similarly Eq.'s (\ref{pi2notinv}) and (\ref{pi2inv}) should now 
include  the corrected $\Pi^{\wedge\vee}_E$. Thus $\Pi_2$ is given by
\begin{eqnarray}
{\Pi}_2 
&=&{g^2N_c\over4\pi^2}
\bigg[\sum_{k=1}^\infty{u^k\over k^2}
\left(4M[\psi(M)+\gamma]-{(M-1)(11M-1)\over3M}
\right)\nonumber\\
& &
-\sum_{k=1}^\infty u^k\sum_{l=1}^{M-1}{lh_k(l)\over Mk}
-\sum_{k=1}^\infty f_k{u^k\over k}
\left(4M[\psi(M)+\gamma]-{7(M-1)\over2}\right)
\bigg],
\label{pi2new}
\end{eqnarray}
and the expansion of Eq.~(\ref{hexpansion}) about $u=1$ should
now be
\begin{eqnarray}
-\sum_{k=1}^\infty u^k\sum_{l=1}^{M-1}
{lh_k(l)\over Mk}&=&-\sum_{k=1}^\infty \sum_{l=1}^{M-1}
{lh_k(l)\over Mk}-(u-1)\sum_{k=1}^\infty \sum_{l=1}^{M-1}
{lh_k(l)\over M}+{\cal O}((u-1)^2)\nonumber\\
&=&-\sum_{k=1}^\infty \sum_{l=1}^{M-1}
{lh_k(l)\over Mk}-(u-1)
{M-1\over 2}+{\cal O}((u-1)^2).
\end{eqnarray}
The constraint equation that replaces (\ref{hsumold}) is
\begin{equation}
\sum_{l=1}^{M-1}{l\over M}\sum_{k=1}^\infty{h_k(l)\over k}
= -{(M-1)(M-2)\over6M}{\pi^2\over6},
\end{equation}
which can be satisfied if we require
\begin{eqnarray}
\sum_{k=1}^\infty{h_k(l)\over k}&=&
-{\pi^2\over18}\left(1-{1\over l}\right).
\label{hsum}
\end{eqnarray}
We prefer this approach to that of a mass counter-term, since it is 
possible that this is a uniform description that works order by order 
(at each order the cancellation places constraints on higher moments
of $h$). The hope is that this can be used non-perturbatively.

In the continuum limit $M(1-u)\to Q^2/2T_0$ stays finite and 
we find that $\Pi_2$ tends to
\begin{eqnarray}
\Pi_2\to{g^2N_c\over16\pi^2}{Q^2\over T_0}\left\{
\left[8(\ln M+\gamma)-{22\over3}\right]\ln{Q^2\over2MT_0}+{4\over3}
\right\}.
\end{eqnarray}
Remembering that our $\Pi_2$ is a factor of $-Q^2/T_0$ times
that defined in  Ref.\ \cite{thornfreedom}, we find agreement
for the coefficient of $Q^2 \ln Q^2$, provided we identify
$Me^\gamma$ with $Q^+/\epsilon$ and $2MT_0$ with $\Lambda^2$. 
We do not get, nor should we expect, the same (finite) coefficient of $Q^2$. 
\subsection{A Brief Calculation of $\Pi$}
In the work just completed, we  deliberately
kept the $A_+$ degree of freedom in the
graphical rules in order to keep the calculation as
close as possible to one in other gauges. However,
having seen how all of the graphs with longitudinal
gluons combine so nicely, it is appropriate to
note that the calculation with
$A_+$ explicitly eliminated, so that the Feynman rules refer only to
the transverse gluons, is much more compact and efficient.
With our prescription we can exploit this simplification
by using the modified cubic vertices Eq.\ (\ref{modcubic}) for
the transverse gluons. At the same time we retain
our replacement of both the bare and induced quartic vertices by
the exchange of two short-lived scalars. The self-energy
diagrams involving those scalars will be exactly as 
described previously (i.e. the terms involving $f$'s and $h$'s). 
However, all of the remaining
contributions to $\Pi^{kl}$ are reduced to the two
or three diagrams involving only transverse gluons
and the modified cubic vertices. 

\begin{figure}[ht]
\begin{center}
\begin{tabular}{|c|c||c|c|}
\hline
&&&\\[-.3cm]
$
\begin{array}[c]{c}
\psfig{file=3VertexUpUpUp.eps,height=0.5in}
\end{array}
$
& 
$0$
&
$
\begin{array}[c]{c}
\psfig{file=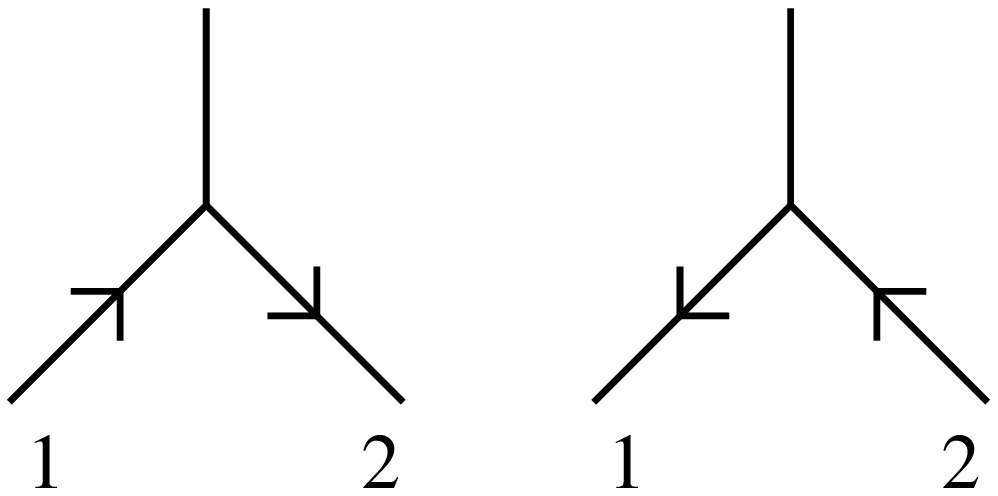,height=0.6in}
\end{array}
$
& 
$ {g\over T_0}(M_2-M_1)$ \\
\hline
&&&\\[-.3cm]
$
\begin{array}[c]{c}
\psfig{file=3VertexUpUpDown.eps,height=0.6in}
\end{array}
$
& 
$ -2{g\over T_0} \left({M_1+M_2\over M_1M_2}\right)
(M_1Q^\wedge_2-M_2Q^\wedge_1) $
&
$
\begin{array}[c]{c}
\psfig{file=3VertexUpDownDash.eps,height=0.6in}
\end{array}
$
& 
$+{g\over T_0}$ \\
\hline
&&&\\[-.3cm]
$
\begin{array}[c]{c}
\psfig{file=3VertexDownDownUp.eps,height=0.6in}
\end{array}
$
& 
$ -2{g\over T_0} \left({M_1+M_2\over M_1M_2}\right)
(M_1Q^\vee_2-M_2Q^\vee_1)$
&
$
\begin{array}[c]{c}
\psfig{file=3VertexDownUpDash.eps,height=0.6in}
\end{array}
$
& 
$ -{g\over T_0}$ \\
\hline
&\multicolumn{3}{c|}{}\\[-.2cm]
$
\begin{array}[c]{c}
\psfig{file=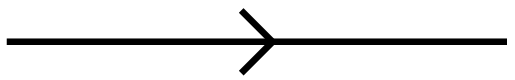,width=0.6in}
\end{array}
$
&
\multicolumn{3}{c|}{
${1 \over 2M} e^{-k{\bf Q}^2/2MT_0}$
} \\[.3cm] 
\hline
&\multicolumn{3}{c|}{}\\[-.2cm]
$
\begin{array}[c]{c}
\psfig{file=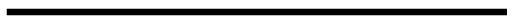,width=0.6in}
\end{array}
$
&
\multicolumn{3}{c|}{
$-f_k {T_0\over M^{2}}e^{-k{\bf Q}^2/2MT_0}$
} \\[.3cm] 
\hline
&\multicolumn{3}{c|}{}\\[-.2cm]
$
\begin{array}[c]{c}
\psfig{file=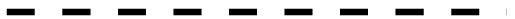,width=0.6in}
\end{array}
$
&
\multicolumn{3}{c|}{
$-h_k T_0 e^{-k{\bf Q}^2/2MT_0}$
} \\[.3cm] 
\hline
\end{tabular}
\end{center}
\caption[]{Summary of discretized Feynman rules using only cubic
vertices. We have explicitly inserted a factor of $1/T_0$ for
each vertex arising from the discretization.}
\label{NewRules}
\end{figure}

Only two diagrams
contribute to  $\Pi^{\wedge\wedge}$, the last two diagrams in
Fig.~\ref{PiUpUp}, and they each involve a prefactor
\begin{eqnarray}
(Mp-lQ)^\wedge(Mp-lQ)^\wedge\to M^2p^\wedge p^\wedge,
\end{eqnarray}
after the shift in momentum. Clearly this integrates to zero so
$\Pi^{\wedge\wedge}=0$.
Finally there are only three
non-tadpole graphs to consider for $\Pi^{\wedge\vee}$, 
namely those in Fig.~\ref{PiUpDownA}. In this case the
relevant prefactors from the three diagrams contributing to 
$\Pi^{\wedge\vee}$ combine as
\begin{eqnarray}
&&2\left[{M^2\over l^2(M-l)^2}+{l^2\over (M-l)^2M^2}+{(M-l)^2\over M^2l^2}
\right](M{\bf p}-l{\bf Q})^2\nonumber\\
&&\hskip1in \qquad\qquad\to2
\left[{M^2\over l^2(M-l)^2}+{l^2\over (M-l)^2M^2}+{(M-l)^2\over M^2l^2}
\right]M^2{\bf p}^2.
\end{eqnarray}
after the usual shift in momentum. Note that because these vertices are 
manifestly Galilei invariant, there is no term proportional to ${\bf Q}^2$
in the prefactor. Thus after integration over loop momentum we are left with
the contribution to $\Pi^{\wedge\vee}$:
\begin{eqnarray}
\Pi^{\wedge\vee}_{\rm non-tadpole}
&=&{g^2N_c\over4\pi^2}{1\over k^2}\sum_{l=1}^{M-1}
\left[{M^2\over l(M-l)}+{l^3\over (M-l)M^2}+{(M-l)^3\over M^2l}
\right]\nonumber\\
&=&{g^2N_c\over4\pi^2}{1\over k^2}\sum_{l=1}^{M-1}
\left[{4M\over l}+2{-3M^2+3Ml-l^2\over M^2}\right]\nonumber\\
&=&{g^2N_c\over4\pi^2}{1\over k^2}
M\left[{4(\psi(M)+\gamma})-{11\over3}+{4\over M}-{1\over3M^2}\right].
\end{eqnarray}
In this way of organizing the calculation, the longitudinal
components of $\Pi^{\mu\nu}$ play no role and the 
new diagrams contributing to $\Pi^{\wedge\vee}$ give 
$\Pi_2$ directly. For the non-tadpole part we find:
\begin{eqnarray}
\Pi_2^{\rm non-tadpole}&=&{g^2N_c\over4\pi^2}\sum_{k=1}^\infty{u^k\over k^2}
M\left[{4(\psi(M)+\gamma})-{11\over3}+{4\over M}
-{1\over3M^2}\right],
\end{eqnarray}
which is clearly a much simpler and more compact calculation! To this
result must be added the contribution of the fictitious scalar
diagrams, which represent the tadpoles. Needless to say,
the calculation of more complicated processes should make
use of these new Feynman rules, which we have summarized in 
Fig.~\ref{NewRules}.

\section{Summing Planar Diagrams: Fishnets}
\label{sec4}
The discretized Feynman rules given at the end of Sec.~\ref{sec3} 
provide a tool to sum classes of diagrams. 
As described in  \cite{thornfishnet} summing diagrams
on a light front has a direct interpretation as the path history quantum
evolution of a system of particles moving in the 
transverse space under Newtonian dynamics. By fixing
the total discretized $P^+=Mm$, the maximum number
of particles present at any time is $M$. Because the
vertices allow particles to fuse and fission, particle
number is not conserved and there is quantum mechanical
mixing between states with any number of particles between
1 and $M$. 

We are particularly interested in the class of planar
diagrams singled out by 't Hooft's $N_c\to\infty$ limit
of QCD. It is actually more precise to think
of this class of diagrams as drawn on a cylinder rather
than a plane: At any time the system of particles is
ordered around a ring, and interactions only exist
between neighbors on this ring. Thus the stage is set for
the particles to bind into a closed polymer chain.
This was previously investigated in \cite{thornfishnet} 
where scalar matrix field
theory with quartic couplings, $\lambda\Tr\phi^4/4$ was
considered. These interactions are attractive (repulsive)
if $\lambda<0$ ($\lambda>0$). Thus bound chains can form
only if $\lambda<0$, the unstable sign. In the interpretation
of the sum of diagrams as a sum over histories of
a system of particles, this sign assures that all histories
contribute with a positive weight. Using the discretization
of $P^+$ and $ix^+$, as reviewed here in Sec.~\ref{sec2},
a strong 't Hooft coupling limit $\lambda N_c\to\infty$ was
formulated and analyzed. This was achieved by focusing
attention on the cylinder diagrams that evolve a system
of particles with $P^+=Mm$ a fixed large number $N$ steps
forward in time. The limit singles out those diagrams in which
every particle has the minimum $P^+=m$ (so the number of
particles is maximal  $=M$), and each propagator
evolves only one time step. For even $M$ these diagrams
include the large fishnet diagrams that form a seamless
web of quartic vertices and propagators, and the 
resulting Gaussian integrals could be identified as
a discretized path integral for a relativistic bosonic string
quantized on the light-cone.

This work on scalar field theory was immediately 
followed by a first attempt to
apply these ideas to large $N_c$ QCD  \cite{browergt}. In
that work QCD was formulated on a light front, with 
$P^+$ and $ix^+$ discretized. The ordinary bare
vertices of QCD, both cubic and quartic, were
used. Because
the quartic coupling in QCD is of order $g^2$, 
the literal strong coupling limit, as formulated
in  \cite{thornfishnet}, favored fishnet
diagrams with {\it only} the primitive (\ie\ non-induced)
quartic couplings.
Actually this conclusion required
the {\it ad hoc} exclusion of the $P^+=0$ exchange
part of the induced quartic interactions arising from
fixing the light-cone gauge. Nonetheless, the
resulting fishnet was very interesting: the spin of
the gluons played the role of the arrows of a certain six
vertex model, known as the F-model~\cite{liebw}. In fact the
four gluon vertices of the field theory were exactly
the vertices of the F-model. The
fact that some of the vertex weights were negative did
not cause problems for the leading strong coupling 
 fishnets because those diagrams always had an {\it even}
number of negative weight vertices. However the problem 
with them reappears at next order
because the deletion of
a single negative vertex reveals a {\it repulsive}
nearest neighbor interaction in that spin channel\footnote{
Contrast this with the cubic scalar theory where
the sign of the coupling is indeed irrelevant because
the deletion of a {\it single} cubic coupling is not
allowed by the Feynman rules.}.
Thus that channel could not have formed a bond in
the first place. In some spin channels there were also positive
weights, so that the bonds {\it could} form. 
However, unfortunately for these fishnets,
the attractive channels are ferromagnetic: the
only long polymers that could be formed by these interactions 
would have enormous spin!

The problem is that at strong coupling only the  quartic 
interactions survived with the discretization of
 \cite{browergt}. The spin-spin interaction from gluon
exchange has anti-ferromagnetic behavior, and it is
possible that a discretization that allowed the exchange 
interaction to compete with the quartic interaction could
cure this problem. To explore this
possibility, one of us examined
the relative strengths of quartic and cubic exchange
interactions for neighbors on a gluonic chain by
putting the two gluons in a spherical MIT bag   \cite{thorngluebag}. 
In that
context one can see explicitly, not only that the
cubic exchange of a transverse gluon is anti-ferromagnetic,
but that its strength (at least in weak coupling perturbation
theory) is more than sufficient to
reverse the ferromagnetic character of the quartic
interaction.  A major shortcoming of the discretization
of   \cite{browergt} is that at strong coupling the
cubic interactions have no opportunity to compete
with the quartic interactions.

\begin{figure}[ht]
\centerline{\psfig{file=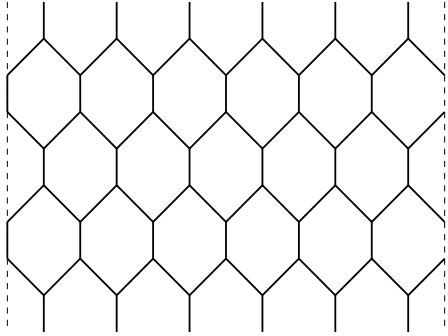,height=1.75in}}
\caption[]{Hexagonal cell fishnet made of only cubic vertices. The
dashed lines indicate the closed cylindrical topology.}
\label{Fishnet}
\end{figure}

The discretization developed in Sec.~\ref{sec2} of this
paper is more promising. In fact, we do away
with quartic interactions completely: All interactions
are cubic! The quartic interactions have been replaced
by the exchange of fictitious scalars, and these 
exchanges are not enhanced by strong coupling over the
exchanges of the ordinary transverse gluons. Without
quartic vertices the basic cells of the densest diagrams are 
no longer square but hexagonal: the fishnet 
looks like a honeycomb (see Fig.~\ref{Fishnet}).
These fishnet diagrams require particles with both 1 and
2 units of $P^+$. Thus the leading fishnet structure is
somewhat looser than in the quartic coupling case. The
inclusion of two different values of $P^+$ in the leading
approximation also sets the stage for the emergence of
a string degree of freedom (provided string states
do form!) corresponding to fluctuations
in $P^+$. Such a degree of freedom is expected to
be described at long worldsheet wavelength
by Polyakov's Liouville field  \cite{polyakov}.

Because all of our vertices are cubic, the paradigm
scalar field theory is now $g\Tr\phi^3/3$ (see \cite{dalleyka} for a
discussion of this model in $1+1$ dimensions). Of course,
the presence of factors of transverse momentum in the QCD cubic
vertices will cause a profound qualitative difference
between gauge and scalar theory. For one thing, the scalar
theory is super-renormalizable with the cubic coupling carrying
dimensions of mass. This means that weak or strong coupling
is determined by the size of the ratio $g^2N_c/\mu^2$, where
$\mu$ is a mass scale relevant to the calculated
physical quantity. For example, in  our strong coupling
considerations, we can simply take $\mu^2=T_0$. Also,
since $\phi^3$ is unbounded at large $\phi$ in one
direction, the theory is ultimately unstable,
although this instability is not evident in weak
coupling perturbation theory if the scalar field has a
non-zero mass. But the 
topology of graphs and momentum flow is the same in both
 scalar field theory and our version of discretized
 QCD, so we shall exploit the scalar paradigm to illustrate such
common features. The absence of spin degrees of freedom
in the scalar theory is a helpful simplification for
at least some issues.

Before turning to details, we comment on the  
ambiguities in our setup contained in the values of the
$f$'s and $h$'s (see Eqs (\ref{effnorm}) and (\ref{hnorm})). 
As shown in our study of the gluon
self-energy in Sec.~\ref{sec3}, weak coupling 
perturbation theory constrains
moments of these quantities (for example, at one loop we
find constraints on $\sum f_k/k$ and $\sum h_k/k$).
Since strong coupling emphasizes short times, we
expect this limit to put constraints on the $f_k$ and $h_k$ for
low values of $k$. Thus we gain complementary or
``dual'' information about the theory as we explore
both limits. It should be stressed here that there is
no ``unique'' discretized theory to associate with
continuum QCD: all sorts of lattice details get washed
out in the continuum limit. Instead our goal is to find
a single lattice model that shows continuum QCD at
weak coupling and string theory at strong coupling.
This dual requirement will hopefully help to determine a
unique theory.

As in  \cite{thornfishnet} we shall consider the sum
of cylinder diagrams that evolve a system with $P^+
=Mm$ forward in time by the amount $T=Na$. For 
fixed initial and final states, according to our
prescription, there are only a finite number of
diagrams that contribute. This defines a definite
model that can be studied as a function of
the bare couplings. If the complete sum could
be done for arbitrary $M$ and $N$, one could then
read off the exact spectrum of the continuum
theory by studying the limit $M,N\to\infty$, with
parameters tuned so that the limit is nontrivial.
This is presumably too ambitious, but in the
next section we shall at least be able to deal nonperturbatively 
with some small values of $M$. One might also
envision studying moderate values of $M$ numerically
on a computer. In the rest of this section we shall
discuss the fishnet diagrams that describe the
infinite coupling limit of our model.

Let us first consider the scalar paradigm. It is
sometimes helpful to define a transfer matrix ${\cal T}$ which
evolves one step forward in time. In order to do this,
start with time continuous and express the exact
time evolution by an amount $a$ in the interaction picture:
\begin{eqnarray}
e^{-a(H_0+V)}=e^{-aH_0}\sum_{n=0}^\infty{(-a)^n\over n!}
\int_0^a dt_1dt_2\cdots dt_n T[V_I(t_1)\cdots V_I(t_n)].
\end{eqnarray}
This expression is exact, and of course it does not correspond
to any discretization. Our discretization is given by
approximating each $V_I(t_i)\approx V_I(0)=V$
and only retaining the term $T[V_I(t_1)\cdots V_I(t_n)]$ when each
$V_I$ acts on a {\it different} subsystem of the particles
present initially. We shall therefore write
the transfer matrix for our discretized theory as
\begin{eqnarray}
{\cal T}=e^{-aH_0}\sum_{n=0}^\infty{(-a)^n\over n!}[V^n],
\label{discretetransfer}
\end{eqnarray}
where we understand $[V^n]=0$ unless each of the $n$ $V$'s acts on
a different subsystem of the particles present.
With this understanding we implement
our discretization rule that every line in a diagram propagate {\it at
least} one step in time. These approximations are strictly valid
for sufficiently small $a$ at fixed coupling parameters. But
we use Eq.~\ref{discretetransfer} to define a discretized
fishnet model at fixed finite $a$, which we intend to
study at all values of the coupling, including $g\to\infty$. 
Although the strong coupling limit at fixed $a$ (as always) takes one
far from the original continuum theory quantitatively, 
we hope that it will lead to a new continuum QCD string theory bearing
qualitative resemblance to real QCD. But there is, of course, no
{\it a priori} guarantee of this outcome.

\begin{figure}[ht]
\centerline{\psfig{file=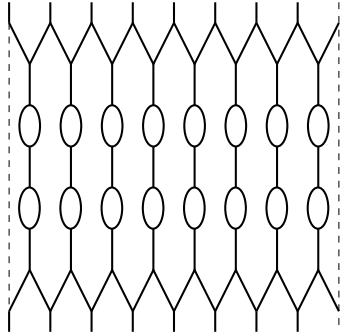,height=1.75in}}
\caption[]{Fishnet with a section of $M=2$ subsystems propagating freely.
The dashed lines again indicate the closed cylindrical topology.}
\label{Bubblenet}
\end{figure}

An efficient way to implement the $N_c\to\infty$ limit is
via the Fock space approach of  \cite{thornfock}. One
chooses a state of the form
\begin{eqnarray}
\ket{\psi}=\sum_{\ell=1}^M{1\over N_c^{\ell/2}}
\sum_{\{M_k\}}\int 
\Tr [a^\dagger_{M_1}({\bf p}_1)\cdots
 a^\dagger_{M_\ell}({\bf p}_\ell)]\ket{0}
\psi_\ell({\bf p}_1,M_1,\ldots,{\bf p}_\ell,M_\ell),
\end{eqnarray}
and applies the transfer matrix keeping only terms that 
survive the $N_c\to\infty$ limit. The second sum is over
all partitions of $M$, such that $M_1+M_2+\cdots +M_\ell = M$.
All such terms retain
the color trace structure and describe interactions
between nearest neighbors as defined by the color trace.
If we wish also to take the infinite 't Hooft coupling
described by the densest fishnet, we choose $M$ even ($M=2n$)
and restrict $\ell$ in the sum to the two values: $\ell=2n$ 
with $M_k=1$ for all $k$ and $\ell=n$ with $M_k=2$ for all $k$, 
and we require that every particle present participate
in an interaction at each time step. This leads to the
following coupled equations for $\psi_{n}$ and $\psi_{2n}$
when $\ket{\psi}$ is an eigenstate of the transfer
matrix:
\begin{eqnarray}
\hskip -.8cm 
t\psi_{2n}({\bf p}_1,\cdots,{\bf p}_{2n})&=&
{g^nN_c^{n/2}e^{-\sum_j({\bf p}_j^2+\mu_0^2)/2T_0}\over(4T_0)^n(2\pi)^{3n/2}}
{1\over2}[\psi_{n}({\bf p}_1+{\bf p}_2, {\bf p}_3+{\bf p}_4,
\cdots, {\bf p}_{2n-1}+{\bf p}_{2n})\nonumber\\
&&\qquad\mbox{} 
+\psi_{n}({\bf p}_2+{\bf p}_3, {\bf p}_4+{\bf p}_5,
\cdots, {\bf p}_{2n}+{\bf p}_1)]\\ 
t\psi_{n}({\bf q}_1,\cdots,{\bf q}_n)&=&
{2g^nN_c^{n/2}e^{-\sum_j({\bf q}_j^2+\mu_0^2)/4T_0}\over(4T_0)^n(2\pi)^{3n/2}}
\int\prod_j{d^2k_j}
\psi_{2n}({\bf k}_1,{\bf q}_1-{\bf k}_1, 
\cdots, {\bf k}_{n},{\bf q}_n-{\bf k}_{n}),
\end{eqnarray}
where $t$ is the eigenvalue of the transfer matrix, and
we have included a bare mass $\mu_0$ for the scalar field. In the 
continuum limit $t\equiv e^{-aE}$. We have suppressed the $M_k$'s 
in the arguments of $\psi_n$ and $\psi_{2n}$ due to their simplicity.
Clearly, we can eliminate $\psi_{2n}$
to obtain a single equation for $\psi_n$:
\begin{eqnarray}
t^2\psi_{n}({\bf q}_1,\cdots,{\bf q}_n)&=&
\left({g^2N_ce^{-5\mu_0^2/4T_0}\over16T_0(2\pi)^{3}}\right)^ne^{-\sum_j{\bf q}_j^2/4T_0}
\int\prod_j{d^2k_j\over T_0}
[\psi_{n}({\bf q}_1, {\bf q}_2,
\cdots, {\bf q}_{n})\nonumber\\
&&\qquad\mbox{} 
+\psi_{n}({\bf q}_1+{\bf k}_2-{\bf k}_1,
\cdots, {\bf q}_{n}+{\bf k}_1-{\bf k}_n)]
e^{-\sum_i[{\bf k}_i^2+({\bf q}_i-{\bf k}_i)^2]/2T_0}.
\end{eqnarray}
We see that the $\mu_0$ dependence is a trivial factor in this
strong coupling equation, so we shall set $\mu_0=0$ in the
following.
The integral of the first term in square brackets is elementary
yielding a factor $\pi^ne^{-\sum_i {\bf q}_i^2/4T_0}$. Defining
$$\lambda\equiv {g^2N_c\over128\pi^2T_0},$$
and rearranging the equation leads to
\begin{eqnarray}
&&\left(t^2-\lambda^ne^{-\sum_j{\bf q}_j^2/2T_0}\right)
\psi_{n}({\bf q}_i)\nonumber\\
&&\qquad\qquad=\lambda^n e^{-\sum_j{\bf q}_j^2/4T_0}
\int\prod_j{d^2k_j\over \pi T_0}
\psi_{n}({\bf q}_i+{\bf k}_{i+1}-{\bf k}_i)
e^{-\sum_i[{\bf k}_i^2+({\bf q}_i-{\bf k}_i)^2]/2T_0}\nonumber\\
&&\qquad\qquad=\lambda^n e^{-\sum_j{\bf q}_j^2/2T_0}
\int\prod_j{d^2k_j\over \pi T_0}
\psi_{n}({\bf q}_i/2+{\bf q}_{i+1}/2+{\bf k}_{i+1}-{\bf k}_i)
e^{-\sum_i{\bf k}_i^2/T_0},
\label{inteqforfn}
\end{eqnarray}
where in the last line we have shifted integration variables
to complete the square in the Gaussian exponent.
This equation sums diagrams including not only the basic fishnet, but
also fishnet diagrams containing any number of 
time intervals in which $n$  subsystems
each with $M=2$ propagate freely for arbitrary lengths
of time (see Fig.~\ref{Bubblenet}). 
This complication is described by
the second term on the l.h.s. In considering the effect of
this term, one should keep in mind that no self mass
counter-terms have been included in the derivation of
(\ref{inteqforfn}). For example, in weak coupling
perturbation theory, the self energy bubble by itself would
lower the scalar mass squared by an infinite amount. To keep the
scalar mass non-tachyonic at weak coupling one needs a mass counter-term.
With discrete time, it is convenient (as in our
treatment of tadpoles) to spread such a mass
counter-term over several time steps by introducing a short
lived fictitious scalar with a quadratic coupling
to the real scalar field of order $g$. 
\begin{figure}[ht]
\vskip .5cm
\centerline{\psfig{file=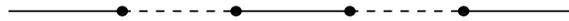,width=3in}}
\vskip .5cm
\caption[]{Additional diagrams introduced by spreading out a 
mass counter-term.}
\label{SolidDashSolid}
\end{figure}
We would then have additional diagrams as in Fig.~\ref{SolidDashSolid}, 
and the upshot for Eq.~(\ref{inteqforfn}) would be an adjustable 
coefficient in front of the second term on the l.h.s.:
\begin{eqnarray}
&&\left(t^2-\lambda^n(1-\delta_1)e^{-\sum_j{\bf q}_j^2/2T_0}\right)
\psi_{n}({\bf q}_i)\nonumber\\
&&\qquad\qquad=\lambda^n e^{-\sum_j{\bf q}_j^2/2T_0}
\int\prod_j{d^2k_j\over \pi T_0}
\psi_{n}({\bf q}_i/2+{\bf q}_{i+1}/2+{\bf k}_{i+1}-{\bf k}_i)
e^{-\sum_i{\bf k}_i^2/T_0}.
\label{inteqforfn2}
\end{eqnarray}
We shall begin by dropping this term completely (\ie by choosing
$\delta_1=1$) 
to determine the contribution of the basic fishnet for
the scalar cubic theory. 
Later, we shall comment on the effect of $\delta_1<1$.

In order to understand the dynamics inherent in the basic
fishnet, it is helpful to express the integral transform
on the r.h.s.\ of (\ref{inteqforfn}) as an operator in the
state space of a first quantized system of $n$ particles.
It is straightforward to show that the appropriate operator
is given by 
\begin{eqnarray}
\Upsilon\equiv\lambda^n e^{-\sum_j{\bf\hat p}_j^2/2T_0}
\Omega e^{-T_0\sum_i({\bf\hat x}_{i-1}
-{\bf\hat x}_{i})^2/4},
\label{fntransfer}
\end{eqnarray}
where the ${\bf\hat p}$'s and ${\bf\hat x}$'s are the momentum and
coordinate operators for the $n$ particle system. Here $\Omega$
is an operator defined in momentum and coordinate bases by
\begin{eqnarray}
\Omega\equiv\int\prod_j d^2p_j\ket{{\bf p}_k}\bra{M_{kl}{\bf p}_l}
\equiv\int\prod_j d^2x_j\ket{{\bf x}_lM_{lk}}\bra{{\bf x}_k},
\end{eqnarray}
and $M$ is an $n\times n$ matrix defined by
\begin{eqnarray}
M_{kl}{\bf p}_l\equiv {{\bf p}_k+{\bf p}_{k+1}\over2}.
\end{eqnarray}
It is easy to check that $M$ is invertible provided $n$ is
odd. For $n$ even there is a zero eigenvalue which must be
separated and handled before continuing the analysis. 
For simplicity, we assume $n$ is
odd in the following discussion.
One can readily verify that $\Omega$ has the following action on
the coordinates and momenta:
\begin{eqnarray}
\Omega {\bf\hat x}_k&=&{\bf\hat x}_l(M^{-1})_{lk}\Omega\\
\Omega {\bf\hat p}_k&=&M_{kl}{\bf\hat p}_l\Omega.
\end{eqnarray}

Because of the Gaussian structure of $\Upsilon$, it also has a
linear action on the coordinates and momenta:
\begin{eqnarray}
\Upsilon {\bf\hat x}_k&=&[{\bf\hat x}_l+i{\bf\hat p}_l/T_0](M^{-1})_{lk}
\Upsilon \label{Upsilonx} \\
\Upsilon {\bf\hat p}_k&=&\left(M_{kl}{\bf\hat p}_l
-i{T_0\over2}[{\bf\hat x}_l+i{\bf\hat p}_l/T_0][2(M^{-1})_{lk}
-(M^{-1})_{l,k+1}-(M^{-1})_{l,k-1}]\right)\Upsilon.
\end{eqnarray}
This linear action can be diagonalized by passing to normal modes:
\begin{eqnarray}
{\bf\tilde x}_l&\equiv&{1\over\sqrt{n}}\sum_k{\bf\hat x}_ke^{-2\pi ilk/n}\\
{\bf\tilde p}_l&\equiv&{1\over\sqrt{n}}\sum_k{\bf\hat p}_ke^{-2\pi ilk/n}.
\end{eqnarray}
One then finds that the modes $l$ all decouple from one another
under the action of $\Upsilon$:
\begin{eqnarray}
\Upsilon {\bf\tilde x}_l&=&{2[{\bf\tilde x}_l+i{\bf\tilde p}_l/T_0]
\over 1+e^{-2\pi il/n}}\Upsilon
=\left[1+i\tan{l\pi\over n}\right][{\bf\tilde x}_l+i{\bf\tilde p}_l/T_0]
\Upsilon\\
\Upsilon {\bf\tilde p}_l&=&\left[{1+e^{2\pi il/n}\over2}{\bf\tilde p}_l
-i\left(1-\cos{2\pi l\over n}\right){2T_0[{\bf\tilde x}_l
+i{\bf\tilde p}_l/T_0]\over 1+e^{-2\pi il/n}}\right]\Upsilon\nonumber\\
&=&\left[1+i\tan{l\pi\over n}\right]\left[
\left(1+\sin^2{l\pi\over n}\right){\bf\tilde p}_l
-2iT_0\sin^2{l\pi\over n}{\bf\tilde x}_l\right]\Upsilon.
\end{eqnarray}

We can now search for eigenoperators of the form ${\bf\tilde x}_l
+\xi_l{\bf\tilde p}_l$. This leads to a quadratic equation for
$\xi_l$,
$$\xi_l^2-{i\over2T_0}\xi_l+{1\over2T_0^2\sin^2(l\pi/n)}=0,$$
with solutions
\begin{eqnarray}
\xi^\pm_l={i\over4T_0}\left(1\pm\sqrt{1+{8\over\sin^2(l\pi/n)}}\right).
\end{eqnarray}
These eigenoperators change the eigenvalue of $\Upsilon$ by a
factor
\begin{eqnarray}
\Lambda_\pm &=&
\left[1+i\tan{l\pi\over n}\right]\left[1-2i\xi^\pm_lT_0
\sin^2{l\pi\over n}\right] \nonumber\\
&=&\left[1+i\tan{l\pi\over n}\right]
\left[1+{1\over2}\sin^2{l\pi\over n}\pm{1\over2}\sin{l\pi\over n}
\sqrt{8+\sin^2{l\pi\over n}}\;\right].
\end{eqnarray}
Note that these eigenvalues are not real because $\Upsilon$ is
not a Hermitian operator. However, also note that the second factor
is positive for both branches and for $0<l<n$. The eigenvalue is
therefore always in the right half complex plane. We also have that 
$|\Lambda_+\Lambda_-|=1$ which implies that 
$|\Lambda_+| > 1 > |\Lambda_-|$. Moreover,
the first factor which contains the complex phase can be rewritten
in two ways
\begin{eqnarray}
1+i\tan{l\pi\over n}
={1\over\cos(l\pi/n)}e^{il\pi/n}=
{1\over\cos((n-l)\pi/n)}e^{-i(n-l)\pi/n},
\end{eqnarray}
which shows that the phase is proportional to the fishnet
momentum created by the eigenoperator: 
$l/n$ for $l<n/2$ and $-(n-l)/n$ for $l>n/2$. Cyclic
symmetry of the initial wavefunction implies that the total
fishnet momentum must be $0$. 

The ground state (belonging to the largest eigenvalue $\Upsilon$) 
is determined by the condition that it be annihilated
by all the eigenoperators which increase $t^2$, 
${\bf\tilde x}_l +\xi_l^+{\bf\tilde p}_l$. 
Its wavefunction is therefore proportional to the Gaussian 
(with normalization ${\cal N}$)
\begin{eqnarray}
\Psi_G = {\cal N} \exp\left\{-\sum_{l=1}^{n-1}{2T_0\sin(l\pi/n)\over
\sin(l\pi/n)+\sqrt{8+\sin^2(l\pi/n)}}{\bf\tilde x}_l\cdot
{\bf\tilde x}_{n-l}\right\},
\label{eigenfunction}
\end{eqnarray}
which is always damped because ${\bf\tilde x}_l^*=
{\bf\tilde x}_{n-l}$. The eigenvalue corresponding to this state 
can be obtained in the following way. First, we observe that
\begin{equation}
\Omega 1 = \int d^2p_k \,\delta(M_{kl}p_l) = {1 \over (\det M)^2} = 
2^{2(n-1)}.
\end{equation}
Then together with Eq.~(\ref{Upsilonx}) we get that
\begin{equation}
\Upsilon \Psi_G = {\cal N} 2^{2(n-1)} {\rm exp}
\left\{{-\sum_{l=1}^{(n-l)/2} {{\bf\tilde p}_l{\bf\tilde p}_{n-l} 
\over T_0}} \right\} {\rm exp} \left\{ 
{-\sum_{l=1}^{(n-l)/2} {{\bf\tilde x}_l{\bf\tilde x}_{n-l} 
\over\cos^2 (l\pi/n)}
\left[{i\over \xi_l^+}+2T_0\sin^2 {l\pi\over n} \right]}\right\}.
\end{equation}
Thus with the use of the identities
\begin{equation}
e^{-\alpha{\bf\tilde p}_l{\bf\tilde p}_{n-l}}
e^{-\beta{\bf\tilde x}_l{\bf\tilde x}_{n-l}}
={1 \over (1+\alpha\beta)^2} e^{-{\beta\over 1+\alpha\beta}
{\bf\tilde x}_l{\bf\tilde x}_{n-l}},
\end{equation}
and 
\begin{equation}
\prod_{l=1}^{(n-1)} \cos^2 (l\pi/n) = {1\over 2^{2(n-1)}},
\end{equation}
we find that the eigenvalue for the ground state is given by
\begin{eqnarray}
t_G^2 = \lambda^{n} \prod_{l=1}^{(n-1)/2}
\left[ 1 + {1\over2} \sin^2 (l\pi/n) + 
{1\over2} \sin (l\pi/n) \sqrt{8+\sin^2 (l\pi/n)}
\right]^{-2}.
\label{eigenvalue}
\end{eqnarray}
Since it is positive, all cyclically symmetric states generated
by applying suitable monomials of the eigenoperators to
the ground state will have positive eigenvalues of $\Upsilon$. 

Clearly the long fishnet
wavelength excitations show behavior identical to
those of the continuous light-cone quantized bosonic string.
The excited states are obtained by applying
appropriate zero momentum monomials of the eigenoperators
${\bf\tilde x}_l+\xi_l^-{\bf\tilde p}_l$ to $\Psi_G$.
From the interpretation $t=e^{-aE}$, we see that 
\begin{eqnarray}
E_G &=& - {1 \over a} \ln t_G \nonumber \\
&=& 
{n \over 2a} \int_0^1 dv \ln\left[ {1\over\lambda}{\left(
1 + {1\over2} \sin^2 {\pi v} + 
{1\over2} \sin {\pi v} \sqrt{8+\sin^2 {\pi v}} \right)}\right]
- {\pi T_0 \over 6\sqrt{2} nm} + {\cal O}\left({1 \over n^2}\right), 
\hskip .5cm
\end{eqnarray}
where we have used the Euler-Maclaurin summation formula for large
$n$
\begin{eqnarray}
{1\over n}\sum_{l=1}^{n-1}F\left({l\over n}\right)=\int_0^1 dv F(v)
-{1\over 2n}[F(0)+F(1)]+\sum_{k\geq1}{B_{2k}\over(2k)!n^{2k}}
[F^{(2k-1)}(1)-F^{(2k-1)}(0)].
\label{EulerMaclaurin}
\end{eqnarray}
In lattice string theory the bulk term proportional to $n$ contains no
physics and can be dropped (see \cite{gilest}). The ground state
string mass squared is predicted to be (recall that $P^+ = 2nm$)
\begin{equation}
M_G^2 = 2P^+E_G = - {\pi T_0 \sqrt2 \over 3}.
\label{StringMass}
\end{equation}
We also see that the basic energy splittings are given for 
$l\ll n$ by
\begin{equation}
\Delta E=\Delta P^-\approx {\pi l\sqrt2\over 2an}
={\pi lT_0\sqrt2\over P^+},
\end{equation}
or splittings in mass squared of
\begin{equation}
\Delta {\cal M}^2=2P^+\Delta P^-=2\pi lT_0\sqrt2.
\end{equation}
This result shows that the string arising from our
basic cubic fishnet has an effective rest tension of $T_0/\sqrt2$, 
corresponding to a Regge slope parameter $\alpha^\prime= 1/\sqrt2\pi T_0$.
Noting that here the transverse dimensionality is 2 and not 24, 
Eq.~(\ref{StringMass}) gives the usual result of bosonic string
theory, $-d/6\alpha^\prime$.

This is all for the basic cubic scalar fishnet, in which the
second term on the l.h.s.\ of Eq.~(\ref{inteqforfn}) is
tuned to zero. Including that term, we find a solution for general $n$
intractable. However, qualitatively, we can say that it
introduces a continuum threshold at $t=\lambda^{n/2}(1-\delta_1)^{n/2}$,
corresponding to a threshold energy $E_{th}=-(n/2a)
\ln(\lambda(1-\delta_1))$.
As long as the basic fishnet described above produces a
ground state energy $E_G<E_{th}$, we can expect qualitatively
similar physics for large $n$. However, for $E_G=E_{th}$ we can expect that
the seamless fishnet structure begins to be disrupted with a dramatic
qualitative change in the physics. We shall explore this effect
for small values of $M$ in the next section.

\begin{figure}[ht]
\centerline{\psfig{file=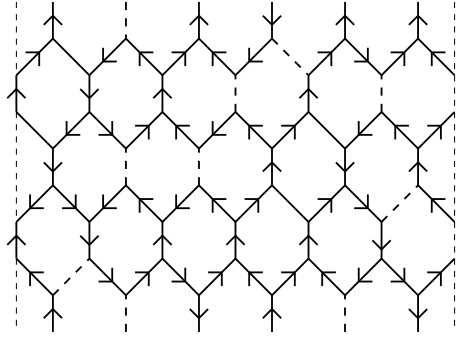,height=1.75in}}
\caption[]{Typical example of a strong coupling QCD fishnet.}
\label{qcdfishnet}
\end{figure}
Finally, let us turn to QCD. The first major
difference is that each line can
exist in four different internal states,  corresponding to
the two polarizations of the transverse gluon (with spin $S^z=\pm1$)
and the two fictitious scalars introduced to simulate the quartic
interactions (see Fig.~\ref{qcdfishnet} for an example of a generic
dense QCD fishnet lattice). In Fock space language this can be
described by affixing a 4-valued index to the creation operators.
The basic fishnet diagrams will be the same as in the 
cubic scalar theory, but with the complication that the
vertex value depends on the states of the lines entering it.
In particular, some of the vertices are linear in the 
transverse momenta of the incoming lines, leading to an
interesting spin-orbit coupling on the fishnet world sheet.
Thus the QCD fishnet dynamics requires the solution of
a two-dimensional lattice spin system with a 
nontrivial interplay between the internal ``spin'' variables
and the structure of the lattice itself. Contrast this with
the fishnet contemplated in \cite{browergt} and based solely
on the quartic coupling. In the latter situation the spin degrees
of freedom decoupled from the transverse coordinate degrees of
freedom and corresponded to the soluble F-model, one of the 
6-vertex models. The fishnet model we are proposing here has a
considerably richer structure, which we shall begin to explore
for small values of $M$ in the next section.

\section{Baby Fishnets}
\label{sec5}
In Section~\ref{sec4} we discussed how our discretized Feynman rules
can be used to determine the dense QCD fishnets for strong
coupling. This was also discussed in the context of the paradigm 
cubic scalar theory. In this section we are interested in the dynamics
of the discretized theory away from strong coupling. Potentially the
strong coupling limit will not be described only by the dense fishnet 
lattice. While the ultimate goal is to do this for
$M\rightarrow\infty$ (remember $P^+=Mm$), we will begin by analyzing
systems with small values of $M$ (i.e. baby fishnets).

The simplest non-trivial QCD fishnet has $M=2$. 
We always understand our fishnets to propagate color singlet
systems so that they have cylindrical topology.
Then the $M=2$ fishnet has no interesting dynamics due to the fact
that color-singlet gluons decouple from gluon bubbles. 
However, the color adjoint $M=2$ gluon propagator which 
plays an important role as a subsystem of larger diagrams,
can be solved to all orders in perturbation theory due its simplicity. 
We leave investigation of $M\ge3$ to the future.
Another possible avenue of investigation is systems
involving sources rather than pure glue,
(such configurations are discussed in \cite{rozowskyt}). 
We also defer exploration of such systems.

In Section~\ref{sec4} we were only able to solve the strong coupling 
cubic scalar fishnet for general $n$ with the term on the l.h.s.\ of
Eq.~(\ref{inteqforfn2}) cancelled via a mass counter-term 
(we will define $\kappa^2=1-\delta_1$). 
But for special cases of $M=2n$ we can solve
(\ref{inteqforfn2}) including $\kappa^2$-term. The simplest of these is
the $M=2$ scalar fishnet, however, in this case the only effect of the
$\kappa^2$-term is to
rescale the solution presented in Section~\ref{sec4}. For the more
interesting cases of $M=4,6$ we will see that they too can be solved.
For $M=6$ we restrict attention to the s-wave sector.
 
\subsection{$M\!=\!2$ States of QCD}
\begin{figure}[ht]
\centerline{\psfig{file=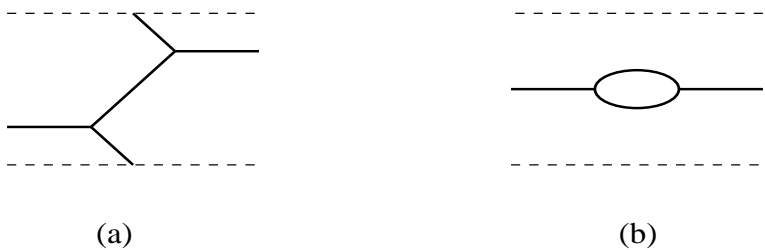,width=4in,height=1.25in}}
\caption[]{Here we see how bubbles of a generic $M\!=\!2$ fishnet lattice 
can close in 2 ways due to the cylindrical topology. The
wraparound diagram (a) has the opposite sign to the bubble
diagram (b), so that they cancel.}
\label{channelalt}
\end{figure}

The $M\!=\!2$ color singlet glueball states display no dynamics, because our
discretization with exclusively cubic vertices only allows interactions
via mixing between 
one gluon and two gluon states, and there is no interacting
color singlet gluon. (Even if the gauge group is $U(N_c)$
the abelian gluon completely decouples in the pure gauge
theory.) To understand this decoupling in terms of our
Feynman rules, note that on a cylinder the gluon self
energy bubble can close in two ways as in  Fig.~\ref{channelalt}.
It is then easy to see from our rules that the two diagrams
are equal in magnitude and opposite in sign. The conclusion is
that the $M\!=\!2$ color singlet channel consists of two
free $M\!=\!1$ gluons or, in the $U(N_c)$ case, a single free
$M\!=\!2$ gluon. This trivial situation is due to the manner
in which the quartic vertices of the initial gauge theory
have been replaced by scalar exchange. Only the ``direct
channel'' scalar exchange is allowed at $M\!=\!2$, and the part of the
quartic vertex that is described by the ``crossed'' channel exchange only
makes its appearance for $M\geq3$.

Thus the only nontrivial $M\!=\!2$ channels are color
non-singlets. Moreover, fixing $M\!=\!2$ limits the allowed 
diagrams so drastically that
the nonvanishing ones can be explicitly summed to all orders
in perturbation theory. We first look at the $M\!=\!2$ gluon
propagator, which can be simply read off from Sec \ref{sec3}. 
For simplicity we work in the center of mass frame. 
\begin{figure}[ht]
\centerline{\psfig{file=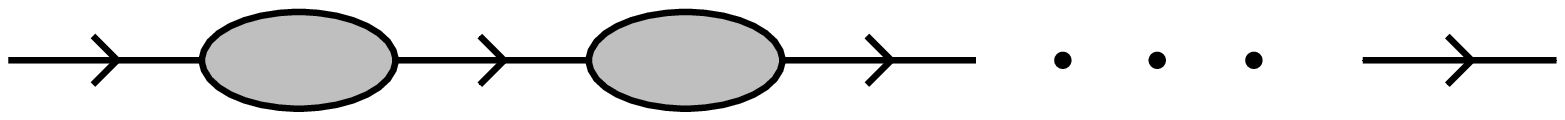,width=5in}}
\caption[]{Gluon propagator for $M\!=\!2$.}
\label{channel1}
\end{figure} 
The diagrams that contribute to the gluon propagator are depicted in 
Fig.~\ref{channel1}. The shaded bubble corresponds to all the one-loop
bubble diagrams that contribute to the transverse gluon self
energy, $\Pi^{\wedge\vee}$ which is obtained
by putting $M\!=\!2$ in Eq.~(\ref{pi2new})
\begin{equation}
\Pi^{\wedge\vee} ={g^2N_c\over8\pi^2}
\bigg[9\sum_{k=1}^\infty{u^k\over k^2}
-9\sum_{k=1}^\infty f_k{u^k\over k}
-\sum_{k=1}^\infty h_k(1){u^k\over k}\bigg],
\label{pi2m2}
\end{equation}
where $u=e^{aE}$ in the center of mass and with 
\begin{equation}
\sum_{k=1}^{\infty} f_k = 1, \qquad 
\sum_{k=1}^{\infty} {f_k\over k} = {\pi^2\over6}, \qquad 
\sum_{k=1}^{\infty} h_k(1) = 1, \qquad 
\sum_{k=1}^{\infty} {h_k(1)\over k} = 0.
\label{m2constraints}
\end{equation}
These constraints on the $f$'s and $h(1)$'s have been determined in 
Section~\ref{sec3} at large $M$ (see (\ref{fsum}) and (\ref{hsum})) 
in order to cancel divergences in $\Pi_2$. 
We tentatively impose the same constraints at all finite $M$ in order
to have a uniform description for all $M$. The exact
transverse gluon propagator for $M\!=\!2$ is (see (\ref{exactprops}))
\begin{eqnarray}
{\tilde D}^{\wedge\vee} &=& 
{u \over 4(1-u) - u\Pi^{\wedge\vee}}\nonumber\\
&=&{u \over 1-u}
\left[4 -{9g^2N_c\over8\pi^2}{u\over (1-u)}\left(
\sum_{k=1}^\infty{u^k\over k^2} 
-\sum_{k=1}^\infty f_k{u^k\over k}
-{1\over9}\sum_{k=1}^\infty h_k(1){u^k\over k}\right)\right]^{-1},
\label{channelone}
\end{eqnarray}
This propagator evolves a spin 1 color adjoint system, which by itself
would not correspond to a glueball, which must be a color singlet. 
Because of its importance for larger diagrams, it is worth
understanding  the energy eigenstates implied by the propagator's pole 
structure. The $(1-u)^{-1}$ factor out front 
is just the massless gluon pole ($E=0$ implies $u=1$). 

Zeroes of the quantity in square brackets in Eq~\ref{channelone}
determine any additional eigenvalues: 
\begin{eqnarray}
{8\pi^2 \over g^2N_c} &=& {9\over4}{u\over (1-u)}\left(
\sum_{k=1}^\infty{u^k\over k^2} 
-\sum_{k=1}^\infty f_k{u^k\over k}
-{1\over9}\sum_{k=1}^\infty h_k(1){u^k\over k}\right).
\end{eqnarray}
Even with the $f$'s and $h(1)$'s general, one can note that the
r.h.s.\ tends to $-\infty$ as $u\to1$, with behavior completely
fixed by the constraints. Also for $u\to0$ the r.h.s.\ vanishes quadratically
as $(9-9f_1-h_1(1))u^2/4$. Therefore if $9-9f_1-h_1(1)>0$
there would be at least two solutions for sufficiently large
$g$, the lowest of which would tend to $u=0$ as $g\to\infty$.
If, however, the inequality were reversed, the r.h.s.\ might
never be positive in which case there would be no solution. 
Alternatively, if it did cross
the axis there would be at least two solutions the
lowest of which would tend to some nonzero $u=u_0>0$ as $g\to\infty$.
It is amusing to see which of these behaviors is suggested
by a minimal solution of the constraints so far imposed (see
(\ref{m2constraints})). For $M\!=\!2$ we can meet the
constraints with only the first two elements of each
series nonzero, which are then fully determined by the constraints:
\begin{equation}
f_1=-1+{\pi^2\over3},\qquad f_2=2-{\pi^2\over3},\qquad
h_1(1)=-1,\qquad h_2(1)=2.
\end{equation}
The eigenvalue equation then reduces to
\begin{equation}
{8\pi^2 \over g^2N_c} = {9\over4}{u\over (1-u)}\left[
{\rm Li}_2(u) - \left(f_1u+f_2{u^2\over2}\right)
-{1\over9}\left(h_1(1)u+h_2(1){u^2\over2}\right)\right],
\label{spinone}
\end{equation}
where ${\rm Li}_2(u)$ is the dilogarithm (Spence) function \cite{lewin}. 
\begin{figure}[ht]
\centerline{\psfig{file=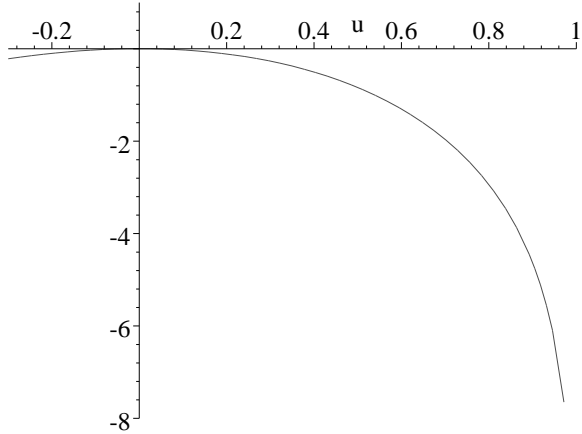,width=3in}}
\caption[]{A plot of Eq.~(\ref{spinone}) with $8\pi^2/g^2N_c$ (along
the vertical axis) against $u$.}
\label{plot1}
\end{figure}
As we can see in Fig.~\ref{plot1} this minimal choice
shows no physical eigenvalue, since there is no positive 
solution for $8\pi^2/g^2N_c$ for any value of $u\in[0,1]$. 

\begin{figure}[ht]
\centerline{\psfig{file=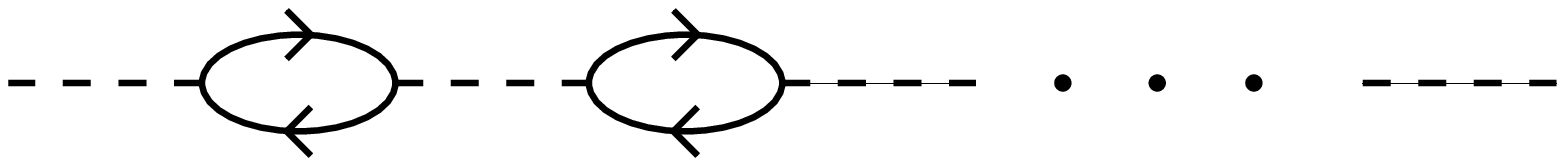,width=5in}}
\caption[]{Magnetic scalar propagator for $M\!=\!2$. In each bubble 
the transverse gluon index may circulate in either direction.}
\label{channel2}
\end{figure}

Finally, the $M\!=\!2$ color adjoint magnetic scalar propagator
also receives self energy corrections which can be summed exactly
(see Fig.~\ref{channel2}). (Note that the
fictitious electric scalar (solid line propagator) does not play
a role in the $M\!=\!2$ channel,
since it's coupling to two $M\!=\!1$ transverse gluons is zero.)
Although the magnetic scalar's contribution to the dynamics of a 
color singlet $M\!=\!2$ glueball
cancels, its propagator describes a spin 0 $M\!=\!2$ color
adjoint subsystem in larger fishnets, and so it is also useful to 
analyze it here.
In these diagrams the bubbles correspond to the one-loop self energy
diagrams of the fictitious magnetic scalar (the dashed scalar). 
\begin{figure}[ht]
\centerline{\psfig{file=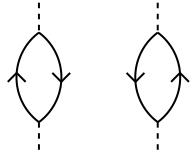,height=0.8in}}
\caption[]{The two bubble diagrams that contribute to the magnetic
scalar self energy, $\Pi_m$.}
\label{PiDashDash}
\end{figure}
The magnetic scalar self energy (see Fig.~\ref{PiDashDash}) is given
by 
\begin{eqnarray}
\Pi_m &=& {g^2N_c \over 2T_0^2} \sum_{k=1}^\infty u^k \int 
{d^2{\bf p}\over (2\pi)^3} e^{-k{\bf p}^2/T_0} \nonumber\\
&=& {g^2N_c\over 16\pi^2}{1\over T_0}\sum_{k=1}^\infty {u^k\over k} 
\nonumber\\
&=& - {g^2N_c\over 16\pi^2}{1\over T_0} \ln(1-u).
\end{eqnarray}
The bare magnetic scalar propagator for $M\!=\!2$ is then
\begin{equation}
D_m = - T_0 \sum_{k=1}^\infty h_k(2) \, u^k,
\end{equation}
where the $h_k(2)$ have to obey the constraints
\begin{equation}
\sum_{k=1}^\infty h_k(2) = 1, \qquad 
\sum_{k=1}^\infty {h_k(2)\over k} = - {\pi^2\over36}.
\end{equation}

The exact propagator is then given by the geometric series
\begin{eqnarray}
{\tilde D}_m &=&  D_m \sum_{l=0}^\infty 
\left( \Pi_m D_m \right)^l 
\:\,=\:\,  {D_m \over 1-\Pi_m D_m}.
\end{eqnarray}
We again investigate possible energy eigenstates by looking at the
pole structure of this amplitude. Focusing on the denominator we see
\begin{eqnarray}
{\tilde D}_m &\propto&  \left[{8\pi^2\over g^2N_c}
-{1\over2}\left(\sum_{k=1}^\infty h_k(2)u^k\right)
\ln(1-u)\right]^{-1}.
\label{channeltwo}
\end{eqnarray}
Again we see the same possible behaviors as in the
case of the gluon propagator (except, of course,
there is no massless pole at $u=1$!). In this case we also 
present the results of choosing a
minimal set of the $h_k(2)$ to satisfy the constraints. Doing this
yields 
\begin{equation}
h_1(2) = -1 - {\pi^2\over18}, \qquad h_2(2) = 2 + {\pi^2\over18}.
\end{equation}
With this set of parameters the denominator factor in
(\ref{channeltwo}) will have a pole if there is a solution to the
following equation
\begin{eqnarray}
{8\pi^2\over g^2N_c} &=& 
{1\over2}\left(h_1(2)u+h_2(2)u^2\right)\ln(1-u) \nonumber\\
&=& - {1\over2} \left|h_1(2)\right|u(1-|\alpha| u)\ln(1-u),
\label{spinzero}
\end{eqnarray}
where $\alpha\equiv h_2(2)/h_1(2)\approx -1.646$.
\begin{figure}[ht]
\centerline{\psfig{file=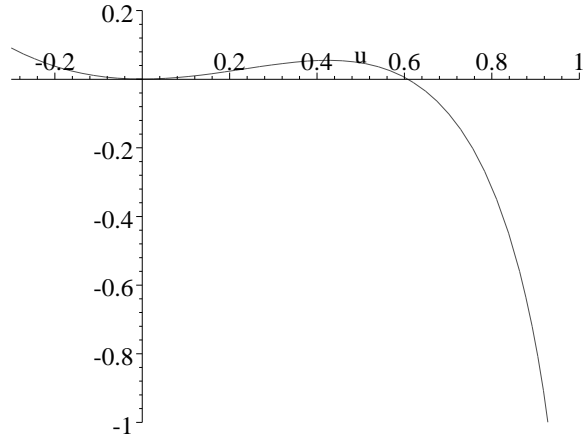,width=3in}}
\caption[]{A plot of Eq.~(\ref{spinzero}) with $8\pi^2/g^2N_c$ (along
the vertical axis) against $u$.}
\label{plot2}
\end{figure}
As we can see in Fig.~\ref{plot2}, this minimal choice shows a physical
bound state (actually two) with $0<u<1$ for the coupling greater than
some critical value, $g^2N_c/8\pi^2\gtrsim18.28$. Such a
bound state would be significant
because it would mean that the short-lived
magnetic scalar, which we have introduced as a device,
can gain longevity
at strong coupling, so that it can play the role of a spin 0 gluon
in larger diagrams. 

There is also a solution for negative $u$ for
all couplings. From the interpretation $u=e^{aE}$ we see that
these solutions would correspond to complex energies with imaginary
part $\pm\pi i/a$. In fact the relation between $u$ and $E$ is
fundamentally ambiguous by the imaginary amount $2\pi i n/a$
simply due to the discretization of $\tau$. We presume that this
ambiguity is simply a lattice artifact, and it seems likely
that the solutions with negative $u$ are also artifacts.
However, the ultimate proof of their artifactual nature must
await a more complete understanding of the continuum limit,
which will involve taking $M\to\infty$ as well as $a\to0$.

\subsection{Strong Coupling $M\!=\!4$ Scalar Fishnet}
Clearly we need to be able to deal with large values of $M$
if we are to understand QCD. In this section, to better
understand what this will involve, we look more closely at the $M\!=\!4$
sector of the simpler paradigm scalar theory. In particular
we would like to explore the effect of 
the $\kappa^2$-term on the l.h.s.\ of the strong coupling 
Eq.~(\ref{inteqforfn2}). Recall that we have set $1-\delta_1=\kappa^2$. 
We dropped this term in the analysis
of Sec.~\ref{sec4} because it made the equation intractable for general $n$. 
However, for the special case of $M\!=\!4$ ($n\equiv M/2=2$) it is
possible to solve this equation.  We also note that 
this case was not covered in Sec.~\ref{sec4} (even with the 
$\kappa^2$-term removed) because of the limitation to odd $n$. The 
additional special case of $M\!=\!6$
will be investigated in the following subsection.

For $M\!=\!4$ the strong coupling eigenvalue equation reads
\begin{eqnarray}
\hskip -1cm 
\left(t^2-\kappa^2\lambda^2e^{-({\bf q}_1^2+{\bf q}_2^2)/2T_0}\right)
\psi_{2}({\bf q}_1,{\bf q}_2)
&=&\nonumber\\
&& \hskip -5cm
\lambda^2 e^{-({\bf q}_1^2+{\bf q}_2^2)/2T_0}
\int{d^2k_1\over \pi T_0}{d^2k_2\over \pi T_0}
\psi_{2}({\bf q}_T/2+{\bf k}_2-{\bf k}_1,
{\bf q}_T/2+{\bf k}_1-{\bf k}_2)e^{-({\bf k}_1^2+{\bf k}_2^2)/T_0},
\end{eqnarray}
where ${\bf q}_T={\bf q}_1+{\bf q}_2$.  After the change of
variables (${\bf k}_T\equiv{\bf k}_1+{\bf k}_2$ and 
${\bf k}\equiv{\bf k}_2-{\bf k}_1$) we can integrate with respect to 
${\bf k}_T$ on the r.h.s.\ What remains is
\begin{eqnarray}
\hskip -1cm 
\left(t^2-\kappa^2\lambda^2e^{-({\bf q}_1^2+{\bf q}_2^2)/2T_0}\right)
\psi_{2}({\bf q}_1,{\bf q}_2)
&=&\nonumber\\
&& \hskip -5cm
{\lambda^2\over2} e^{-({\bf q}_1^2+{\bf q}_2^2)/2T_0}
\int{d^2k\over \pi T_0}
\psi_{2}({\bf q}_T/2+{\bf k},
{\bf q}_T/2-{\bf k})e^{-{\bf k}^2/2T_0}.
\end{eqnarray}

If we now perform a variable transformation on the ${\bf q}$'s 
(${\bf q}_T\equiv{\bf q}_1+{\bf q}_2$ and 
${\bf q}\equiv({\bf q}_1-{\bf q}_2)/2$) then the equation becomes
\begin{eqnarray}
\hskip -1cm 
\left(t^2-\kappa^2\lambda^2e^{-({\bf q}_T^2+4{\bf q}^2)/4T_0}\right)
\psi_{2}({\bf q}_T/2+{\bf q},{\bf q}_T/2-{\bf q})
&=&\nonumber\\
&& \hskip -5cm
{\lambda^2\over2} e^{-({\bf q}_T^2+4{\bf q}^2)/4T_0}
\int{d^2k\over \pi T_0}
\psi_{2}({\bf q}_T/2+{\bf k},
{\bf q}_T/2-{\bf k})e^{-{\bf k}^2/2T_0}.
\label{M4equation}
\end{eqnarray}
We can absorb the ${\bf q}_T$ dependence by scaling 
$t=t_0\exp(-{\bf q}_T^2/8T_0)$, in effect going to the
center of mass frame. 

For the case of ${t}_0^2>\kappa^2\lambda^2$, we can eliminate the $\psi_2$
dependence by manipulating this equation and integrating with respect
to ${d^2q}$. The result
\begin{eqnarray}
1={\lambda^2\over2}\int{d^2q\over\pi T_0}{e^{-3{\bf q}^2/2T_0}
\over t_0^2-\kappa^2\lambda^2e^{-{\bf q}^2/T_0}},
\end{eqnarray}
is a transcendental equation for the eigenvalue, as is readily
seen by direct evaluation of the integral:
\begin{equation}
\kappa^2 = {\hat{t}\over2\kappa}\ln{\hat{t}+\kappa\over\hat{t}-\kappa}-1,
\label{simpleeq}
\end{equation}
where $\hat{t}\equiv t_0/\lambda$ and we have assumed $\kappa^2>0$. 
If $\kappa^2<0$ we write $\kappa\equiv i\alpha$ and the equation becomes
\begin{equation}
1-\alpha^2 = {\hat{t}\over\alpha}\tan^{-1}\left({\alpha\over{\hat t}}\right).
\label{simpleeq2}
\end{equation}
It is immediately clear that a solution exists in this case only for
$\alpha^2=-\kappa^2<1$.

If we analyze (\ref{simpleeq}) we see
that by varying $\hat{t}$ between $\kappa$ and $\infty$ the r.h.s.\ takes
on values between $0$ and $\infty$, thus there is a solution to this
equation for any value of $\kappa$. For the special case of
$\kappa=0$, a more careful analysis of this equation also yields a
solution. For a given eigenvalue solution to (\ref{simpleeq}) the
eigenfunction is given by
\begin{eqnarray}
\psi_{2}({\bf q}) =
{e^{-{\bf q}^2/T_0} \over 2\left({\hat{t}}^2-\kappa^2e^{-{\bf q}^2/T_0}\right)}
\int{d^2k\over \pi T_0}
\psi_{2}({\bf k})e^{-{\bf k}^2/2T_0},
\end{eqnarray}
where the integral on the r.h.s.\ represents a number that can be fixed by
normalization. 

If we refer back to (\ref{M4equation}) we see that if
$\hat{t}^2<\kappa^2$, then a delta function term restricting
momentum to energy shell may be added. For $s$-waves (non
$s$-waves are free), we then have
\begin{eqnarray}
\psi_{2}({\bf q}) = 
A\delta\!\left({\bf q}^2 - T_0\ln{\kappa^2\over\hat{t}^2}\right) +
{e^{-{\bf q}^2/T_0} \over 
2\left({\hat{t}}^2-\kappa^2e^{-{\bf q}^2/T_0}+i\epsilon\right)}
\int{d^2k\over \pi T_0}
\psi_{2}({\bf k})e^{-{\bf k}^2/2T_0},
\end{eqnarray}
where the coefficient of the delta function can be fixed via the
following equation which relates $A$ to the normalization of the
wavefunction,
\begin{eqnarray}
\int{d^2q\over \pi T_0}\psi_{2}({\bf q})e^{-{\bf q}^2/2T_0} = 
{A\over T_0}\left|{\hat{t}\over\kappa}\right| +
\int{d^2q\over \pi T_0}{e^{-3{\bf q}^2/2T_0} \over 
2\left({\hat{t}}^2-\kappa^2e^{-{\bf q}^2/T_0}+i\epsilon\right)}
\int{d^2k\over \pi T_0} \psi_{2}({\bf k})e^{-{\bf k}^2/2T_0}.
\end{eqnarray}
In this case, $\hat{t}^2<\kappa^2$, there is no restriction on the value
of $\hat{t}$, thus this solution corresponds to the continuum.
To summarize the spectrum of $t_0^2$ includes a discrete $s$-wave bound state
and a continuum for $t_0^2<\kappa^2\lambda^2$. Clearly the discrete
state will not change drastically in the limit $\kappa\to0$, but the
continuum would be dramatically squeezed to a set of measure zero
in this limit.

\subsection{Strong Coupling $M\!=\!6$ Scalar Fishnet}
The starting point for evaluating the strong coupling $M\!=\!6$ bit scalar fishnet
diagrams is Eq.~(\ref{inteqforfn2}). For $n=M/2=3$ the equation to solve
is 
\begin{eqnarray}
\hskip -1cm 
\left(t^2-\kappa^2\lambda^3e^{-({\bf q}_1^2+{\bf q}_2^2+{\bf q}_3^2)/2T_0}
\right)\psi_{3}({\bf q}_1,{\bf q}_2,{\bf q}_3)
&=&
\lambda^3 e^{-({\bf q}_{1}^2+{\bf q}_2^2+{\bf q}_3^2)/2T_0}
\int{d^2k_1\over \pi T_0}{d^2k_2\over \pi T_0}{d^2k_3\over \pi T_0}
\nonumber\\
&& \hskip -5.5cm
\psi_{3}(({\bf q}_{1}+{\bf q}_{2})/2+{\bf k}_{21},
({\bf q}_{2}+{\bf q}_{3})/2+{\bf k}_{32},
({\bf q}_{3}+{\bf q}_{1})/2+{\bf k}_{13})
e^{-({\bf k}_1^2+{\bf k}_2^2+{\bf k}_3^2)/T_0},
\end{eqnarray}
where we have used 
${\bf k}_{ij}={\bf k}_i-{\bf k}_j$. By working in the
center of mass frame, we can replace ${\bf q}_3\rightarrow 
-{\bf q}_1-{\bf q}_2$ and the equation becomes
\begin{eqnarray}
\hskip -1cm 
\left(t^2-\kappa^2\lambda^3
e^{-({\bf q}_1^2+{\bf q}_2^2+{\bf q}_1\cdot{\bf q}_2)/T_0}
\right)\psi_{3}({\bf q}_1,{\bf q}_2,-{\bf q}_1-{\bf q}_2)
&=&
\lambda^3 e^{-({\bf q}_{1}^2+{\bf q}_2^2+{\bf q}_1\cdot{\bf q}_2)/T_0}
\int{d^2k_1\over \pi T_0}{d^2k_2\over \pi T_0}{d^2k_3\over \pi T_0}
\nonumber\\
&& \hskip -5cm
\psi_{3}({\bf q}_{1}/2+{\bf q}_{2}/2+{\bf k}_{21},
-{\bf q}_{1}/2+{\bf k}_{32},
-{\bf q}_{2}/2+{\bf k}_{13})
e^{-({\bf k}_1^2+{\bf k}_2^2+{\bf k}_3^2)/T_0}.
\end{eqnarray}
If we change the integration variables ${\bf k}_2$ and ${\bf k}_3$ on
the r.h.s.\ to  
\begin{equation}
{\bf p}_1 = {\bf q}_{1}/2+{\bf q}_{2}/2+{\bf k}_{21}, \qquad
{\bf p}_2 = -{\bf q}_{1}/2+{\bf k}_{32},
\end{equation}
then we see that $\psi_3$ in the integrand on the r.h.s.\ is independent
of ${\bf k}_1$, so the Gaussian integral over ${\bf k}_1$
may be trivially performed by completing the square.

Once this has been done the result is
\begin{eqnarray}
\hskip -1cm 
\left(t^2-\kappa^2\lambda^3
e^{-({\bf q}_1^2+{\bf q}_2^2+{\bf q}_1\cdot{\bf q}_2)/T_0}
\right)\psi_{3}({\bf q}_1,{\bf q}_2)
&=&
{\lambda^3\over3} 
e^{-({\bf q}_{1}^2+{\bf q}_2^2+{\bf q}_1\cdot{\bf q}_2)/T_0}
\int{d^2p_1\over \pi T_0}{d^2p_2\over \pi T_0}
\psi_{3}({\bf p}_{1},{\bf p}_{2}) \nonumber\\
&&\hskip -2.5cm
\times \exp\left[
-{2\over3T_0}\left({\bf p}_1^2+{\bf p}_2^2+{\bf p}_1\cdot{\bf p}_2\right)
-{1\over6T_0}\left({\bf q}_1^2+{\bf q}_2^2+{\bf q}_1\cdot{\bf q}_2\right)
\right. \nonumber \\
&& \hskip -2.5cm
\phantom{\times \exp\left[\right.}\left.
+{1\over3T_0}\left({\bf p}_1\cdot{\bf q}_1+2{\bf p}_1\cdot{\bf q}_2
-{\bf p}_2\cdot{\bf q}_1+{\bf p}_2\cdot{\bf q}_2 \right)\right].
\end{eqnarray}
In this equation we see that except for the last term in the
exponential on the r.h.s.\ this integral equation only depends on the
scalar quantity, ${\bf q}_1^2+{\bf q}_2^2+{\bf q}_1\cdot{\bf
q}_2$. This is not surprising since this scalar quantity is
proportional to ${\bf q}_1^2+{\bf q}_2^2+{\bf q}_3^2$ and is the only
cyclically invariant (${\bf q}_i\rightarrow{\bf q}_{i+1}$) scalar 
of order ${\bf q}^2$. We next perform another change of variables,
\begin{equation}
{{\bf q}_1\over\sqrt{T_0}} = {\bf u}_1 + {\bf u}_2, \qquad
{{\bf q}_2\over\sqrt{T_0}} = {\bf u}_1 - {\bf u}_2, \qquad
{{\bf p}_1\over\sqrt{T_0}} = {\bf v}_1 + {\bf v}_2, \qquad
{{\bf p}_2\over\sqrt{T_0}} = {\bf v}_1 - {\bf v}_2,
\end{equation}
remembering that ${\bf q}_i$, ${\bf p}_i$ and now ${\bf u}_i$, 
${\bf v}_i$ are Euclidean 2-vectors. With these substitutions our
equation becomes
\begin{eqnarray}
\hskip -1cm 
\left(t^2-\kappa^2\lambda^3
e^{-(3{\bf u}_1^2+{\bf u}_2^2)}\right)\psi_{3}({\bf u}_1,{\bf u}_2)
&=&
{4\lambda^3\over3\pi^2} 
e^{-(3{\bf u}_{1}^2+{\bf u}_2^2)}
\int{d^2v_1}{d^2v_2}\,
\psi_{3}({\bf v}_{1},{\bf v}_{2}) \nonumber\\
&&\hskip -6cm
\times \exp\left[
-{2\over3}\left(3{\bf v}_1^2+{\bf v}_2^2\right)
-{1\over6}\left(3{\bf u}_1^2+{\bf u}_2^2\right)
+{\bf u}_1\cdot{\bf v}_1+{\bf u}_1\cdot{\bf v}_2
-{\bf u}_2\cdot{\bf v}_1+{1\over3}{\bf u}_2\cdot{\bf v}_2 \right].
\end{eqnarray}
If we rescale ${\bf u}_2$ and ${\bf v}_2$ by a factor of $\sqrt{3}$,
then we can make the $O(4)$ symmetry manifest (up to the last four
terms in the exponential on the r.h.s.) by combining ${\bf u}_1$ and  
${\bf u}_2$ (${\bf v}_1$ and ${\bf v}_2$) into a Euclidean 4-vector 
${\bf U}$ (${\bf V}$). Thus the equation may
be written as
\begin{eqnarray}
\left(t^2-\kappa^2\lambda^3 e^{-3{\bf U}^2}\right)\psi_{3}({\bf U}) 
&=& {4\lambda^3\over\pi^2} e^{-7{\bf U}^2/2}
\int{d^4V}\, \psi_{3}({\bf V}) 
e^{-2{\bf V}^2+2{\bf V}^{\rm T}\mbox{${\cal R}$}{\bf U}}, \nonumber \\
&=& {4\lambda^3\over\pi^2} e^{-7{\bf U}^2/2}
\int{d^4V}\, \psi_{3}(\mbox{${\cal R}$}{\bf V}) 
e^{-2{\bf V}^2+2{\bf V}\cdot{\bf U}},
\end{eqnarray}
where ${\cal R}$ is the real orthogonal $O(4)$ rotation:
\begin{equation}
\mbox{${\cal R}$} \equiv 
{1\over2} \left(
\begin{array}{cccc}
1&0&-\sqrt{3}&0 \\
0&1&0&-\sqrt{3} \\
\sqrt{3}&0&1&0  \\
0&\sqrt{3}&0&1  \\
\end{array}
\right).
\end{equation}
We note that ${\cal R}^3=-1$. We can search for an
$O(4)$ invariant solution to this equation which is
a function only of the  length, $\|{\bf U}\|$. Although this will not 
yield the most general eigenstate, it is expected to
include the ground state. Plugging in this {\it ansatz},
the equation simplifies to
\begin{eqnarray}
\left(t^2-\kappa^2\lambda^3 e^{-3U^2}\right)\psi_{3}(U) 
&=& {4\lambda^3\over\pi^2} e^{-7U^2/2}
\int{V^3dV}\, \psi_{3}(V) 
e^{-2V^2}\int{d\Omega_{\bf V}}
e^{2{\bf V}\cdot{\bf U}},
\end{eqnarray}
where $U\equiv\|{\bf U}\|$ and $V\equiv\|{\bf V}\|$.
The angular integral on the r.h.s.\ may be evaluated with standard
techniques,
\begin{eqnarray}
\int{d\Omega_{\bf V}}e^{2{\bf V}\cdot{\bf U}} &=&
4\pi\int_{-1}^1d\zeta\sqrt{1-\zeta^2}e^{2UV\zeta} \nonumber \\
&=& {2\pi^2 I_1(2UV)\over UV},
\end{eqnarray} 
where $I_1(x)$ is the modified Bessel function regular at $x=0$.

Thus the integral equation to solve is
\begin{eqnarray}
t^2 \psi_{3}(U) 
&=&\kappa^2\lambda^3 e^{-3U^2}\psi_{3}(U) 
+ {8\lambda^3} {e^{-7U^2/2}\over U}
\int{V^2dV}\,I_1(2UV) \psi_{3}(V) e^{-2V^2}.
\label{iterate}
\end{eqnarray}
The first thing to note is that for $\kappa^2=0$ the eigenfunction 
solution is a Gaussian of the form
\begin{eqnarray}
\psi_{3}(U) = e^{-\xi U^2}, \qquad\mbox{where}\qquad 
\xi={3+\sqrt{105} \over 4}.
\end{eqnarray}
For this solution the corresponding eigenvalue is
\begin{eqnarray}
t^2 = {64\lambda^3\over \left(11+\sqrt{105}\right)^2}.
\end{eqnarray}
Both of these match the values predicted by Equations
(\ref{eigenfunction}) and (\ref{eigenvalue}) for $n=3$. For $\kappa^2$
away from zero we can solve this integral equation by means of an
iterative procedure. Starting with the solution for $\kappa^2=0$ we can
iterate the r.h.s.\ of (\ref{iterate}) repeatedly. This is a convenient
way of solving this equation since the functions generated by the
integral on the r.h.s.\ are always
Gaussian. Thus at each iteration step the solution will be of the form
\begin{equation}
\psi_3(U) = \sum_n c_n e^{-\alpha_n U^2},
\end{equation}
with the number of terms in the sum doubling after each iteration.
For a solution of this form it can be shown that the corresponding
eigenvalue, $t^2$, is given by
\begin{equation}
t^2 = \kappa^2 + {4\sum_n c_n/(2+\alpha_n)^2\over\sum_n c_n}.
\end{equation}

We have tested this iteration procedure numerically and for values of
$\kappa^2$ small ($\kappa^2<0.3$) we see that the wavefunction(eigenvalue)
converges to a well-defined function(value). For $\kappa^2$ larger
(closer to 1) this becomes murkier as one needs a lot more iterations
for the convergence to a value distinct from $\kappa^2$ to be
evident. Another interesting phenomenon is that for
$\kappa^2\lesssim -0.11$, then $t^2$ becomes negative indicating that
no physical solution exists.

\section{Conclusion}
\label{sec6}
In this article, we have refined and extended an approach, 
proposed in the late seventies, to
obtain the large $N_c$ limit of QCD by directly
summing the planar diagrams which survive.
The basic tool is to
define the planar diagrams using light-front space-time
coordinates for which $ix^+$ and the $p^+$ carried by each
gluon are discretized.
This effectively digitizes the sum of diagrams, a first step
toward a numerical evaluation. It also regulates
the usual divergences of Feynman diagrams. We identified several
shortcomings of the discretized model of QCD attempted
in \cite{browergt}, and we proposed an improved formulation which
at least mitigates, and might well overcome, these defects.

Discretization enables a formal strong 't Hooft coupling limit 
of the sum of diagrams. A major disadvantage of the 
discretization of \cite{browergt} was that this formal
limit suppressed the cubic gluonic interaction essential
for the ``anti-ferromagnetic'' ordering of glueball mass
levels: the dominant quartic interaction ordered
levels ferromagnetically. Our new discretized model 
replaces the quartic interactions by the exchange of two kinds of
fictitious ``short-lived'' scalars, so that all interactions can compete
on an equal footing in the strong coupling limit.
The ambiguities inherent in such a replacement can also
be exploited to remove unwanted symmetry violations
induced by the usual ultraviolet divergences present in the
continuum limit.

Having defined our discretized model, we explored its
physical properties in several ways. We first studied the
nature of weak coupling perturbation theory by calculating
the gluon self energy to one loop order, regaining the
known continuum answer. This calculation showed how the
discretization regulates ultraviolet divergences, and how
the ambiguities in the model begin to be fixed by the 
restoration of Poincar\'e invariance. Although we have
not done a two loop calculation, there is sufficient
flexibility in these ambiguities to hope to achieve Poincar\'e
invariance to all orders in perturbation theory. The discretized
model can also be studied in the strong coupling limit,
but in this article we just began this study for QCD by
looking only at states with very small total $P^+=Mm$ for $M\!=\!2$,
where the dynamics is so drastically simplified that
it can be solved exactly. We defer to a future publication
studies of QCD at $M\!=\!3$ and higher. The
continuum limit, of course, will require $M\to\infty$.

As a warmup for going to larger values of $M$, we 
evaluated the strong coupling limit in a paradigm
matrix scalar field theory with only cubic interactions. 
Not surprisingly, the bosonic light-cone string was
obtained. Although
this paradigm model yielded some useful insights into
the nature of large planar diagrams, we stressed that
the corresponding QCD calculation will have profound differences:
for one thing the gluons carry spin, and for another
their interactions show both repulsion and attraction 
depending on the quantum numbers of the channel.
In contrast, the interactions of the scalar theory are
exclusively attractive. Because of this, the strong coupling limit
forced the $p^+$ carried by each scalar quantum to be minimal, \ie
one discretized unit $m$.
This circumstance prevented a ``Liouville'' degree of freedom,
associated with collective fluctuations of the $P^+$ distribution
among the scalar quanta, from arising. Thus the limit must be interpreted
as a critical string theory.

The diversity of interaction signs
of QCD will obviously complicate this outcome. It is possible,
and a major focus for future study, that cancellations
deemphasize the contributions where all quanta carry the minimum
$p^+$ to such a degree that a collective Liouville field 
emerges. Then the strong coupling limit might be a subcritical
version of one of the existing string models.
If so, the Liouville world-sheet field could be
thought of as a fifth dimension, and the dual description
of our model as a field theory at weak coupling and a subcritical string
theory at strong coupling would resemble the anti de Sitter gravity
(AdS) / conformal field theory (CFT) duality of 
\cite{maldacena,gubserkp,wittenholog}.
Another logical possibility, though, is that the strong
coupling limit of large $N_c$ QCD is actually a
novel critical string theory with critical dimension 4.
Of course, it could also turn out that the attempt to reach a reasonable
Poincar\'e invariant strong coupling limit of large $N_c$ QCD simply fails.
After all, continuum QCD is, strictly speaking, {\it not} an infinite coupling
theory in any sense of the word. The coupling is scale
dependent and corresponds to no tunable parameter at all.
The strong coupling limit, as everyone knows, describes
the discretized model and can vary wildly from one discretization
to another. 

Much has been said about the ``holographic'' nature of the
duality mentioned above. We would like to conclude with a
few comments about this. The hologram metaphor was invented
by 't Hooft \cite{thoofthol} to describe a possible 
resolution of the ``information
loss paradox'' of quantum black holes. Since the horizon of
a black hole is two dimensional, it should be possible to
describe all of three dimensional physics by a two dimensional
quantum theory. The discretized model we have presented here is not 
holographic in this sense. The transverse space of
a light front is indeed two dimensional, but the third
longitudinal dimension has not been eliminated: it is present
in the disguised form of a variable Newtonian mass $Mm$ for each
gluon. However, the model is holographic in the higher dimensional
sense described by Witten \cite{wittenholog}. The ``fundamental''
discretized model is $3+1$ dimensional, 2 transverse dimensions,
variable $p^+$ and $x^+$. However in the strong coupling limit
we expect $4+1$ dimensions: the $x^-$ of light-cone string
should emerge as a function of the transverse and Liouville
degrees of freedom. Holography in 't Hooft's sense would 
require a more profound circumstance: there should
be no Liouville field and the variable $p^+$ of
each gluon must itself be a mere collective effect. For example,
the gluon with $M$ units of $p^+$ might be thought of as a
bound system of $M$ minimal $p^+$ ``bits'' \cite{thornfront}.
In that case, the model presented here would just be a stepping
stone toward that more fundamental theory.

\vskip .5cm
 
\underline{Acknowledgments:} We would like to thank M. Brisudova
for helpful criticism of the  manuscript. JSR and CBT would
also like to acknowledge the Aspen Center for Physics where part of
this work was completed. This work was supported in part by the Department
of Energy under Grant No. DE-FG02-97ER-41029.

\setcounter{equation}0
\setcounter{section}0
\renewcommand{\thesection}{\Alph{section}}
\renewcommand{\theequation}{\Alph{section}.\arabic{equation}}
\section{Appendix: Alternate Discretization}
\label{secA}
In this appendix we explore the ramifications of 
the alternate discretization Eq.~(\ref{d--2})
of $D^{--}$. The bare propagator in energy representation is
\begin{eqnarray}
{\tilde D}_{0}^{--}({\bf Q}, M, E)&=&{T_0(e^{aE}-1)\over M^2}{u\over
1-u}.
\end{eqnarray}
The self-energy parts $\Pi^{kl}, \Pi^{k+}$ and $\Pi^{++}$
will of course have different values in this
discretization, but the decompositions (\ref{defpi12}) remain
valid. 
Under the assumption that $\Pi^{\prime\prime}_1
=\Pi^\prime_1=\Pi_1$, the relations of the
exact propagators to the $\Pi$'s are identical to
(\ref{exactprops}) except for ${\tilde D}^{--}$ for which
the relation is
\begin{eqnarray}
{\tilde D}^{--}&=&{T_0\over M^2}\left[{u(e^{aE}-1-{\bf Q}^2/2MT_0)
\over 1-u-T_0u\Pi_1(e^{aE}-1-{\bf Q}^2/2MT_0)}
+{{\bf Q}^2\over2MT_0}{u\over 1-u-u\Pi_{2I}/2M}\right].
\end{eqnarray}

The only parts of the one loop self energy calculation
affected by the different discretization are 
the two diagrams, Fig.~\ref{PiUpDownD}, which have a $D^{--}$
propagator as one of the internal lines. The evaluation
is quite different for this discretization  because the
completion of squares in the second term of Eq.~(\ref{d--2}) leads to
different factors than the first term. The contribution of the first
term involves the exponent
\begin{eqnarray}
{k\over2T_0}\left({{\bf p}^2\over l}+{({\bf Q}-{\bf p})^2
\over M-l}\right)={k\over2T_0}\left({M({\bf p}-l{\bf Q}/M)^2\over l(M-l)}
{}+{{\bf Q}^2
\over M}\right),
\end{eqnarray}
whereas the contribution of the second term involves
\begin{eqnarray}
{(k-1){\bf p}^2\over2T_0 l}+{k({\bf Q}-{\bf p})^2
\over2T_0 (M-l)}={1\over2T_0}\left({M(k-1)+l\over l(M-l)}
{\left({\bf p}-{lk{\bf Q}\over M(k-1)+l}\right)^2}+{k(k-1){\bf Q}^2
\over M(k-1)+l}\right),
\end{eqnarray}
for one of the two diagrams, and for the other a similar
expression with ${\bf p}\to{\bf Q}-{\bf p}$ and $l\to M-l$.
Thus the two contribute equally to $\Pi^{\wedge\vee}_{{\rm D}I}$:
\begin{eqnarray}
\Pi^{\wedge\vee}_{{\rm D}}  
&=&{g^2N_c\over4\pi^2}\sum_{l=1}^{M-1}{(2M-l)^2\over l}e^{-k{\bf
Q}^2/2MT_0}
\left[{e^{kl{\bf Q}^2/2MT_0(M(k-1)+l)}\over
M(k-1)+l}-{1\over Mk}\right], 
\end{eqnarray}
for $k>1$ and by
\begin{eqnarray}
\Pi^{\wedge\vee}_{{\rm D}}&=&
-{g^2N_c\over4\pi^2}\sum_{l=1}^{M-1}{(2M-l)^2\over l}
{e^{-{\bf Q}^2/2MT_0}\over M}, 
\end{eqnarray}
for $k=1$.
Translating to energy representation gives the more compact
\begin{eqnarray}
{\tilde \Pi}^{\wedge\vee}_{\rm D} 
&=&
{g^2N_c\over4\pi^2}
\sum_{l=1}^{M-1}{(2M-l)^2\over l}\sum_{k=1}^\infty
u^k
\left[u{e^{l(k+1){\bf Q}^2/2MT_0(Mk+l)}\over
Mk+l}-{1\over Mk}\right].
\end{eqnarray}
Adding the result of the unchanged first thirteen
diagrams to this and subtracting ${\bf Q}^2\Pi_1/2$ gives $\Pi_2$
for this discretization:
\begin{eqnarray}
{\Pi}_{2}&=&{g^2N_c\over24\pi^2}
{13M^2-12M-1\over M}\sum_{k=1}^\infty{u^k\over k^2}+{g^2N_c\over48\pi^2}
\left(14-{15\over M}
+{1\over M^2}\right){{\bf Q}^2\over T_0}\ln(1-u)\nonumber\\
&&\qquad+{g^2N_c\over4\pi^2}
\sum_{l=1}^{M-1}{(2M-l)^2\over l}\sum_{k=1}^\infty
u^k
\left[u{e^{l(k+1){\bf Q}^2/2MT_0(Mk+l)}\over
Mk+l}-{1\over Mk}\right].
\label{pi2I}
\end{eqnarray}
The violation of Galilei invariance caused by this
alternate discretization 
is apparent from the non-polynomial dependence on ${\bf Q}^2$. However,
one can easily see that each power of ${\bf Q}^2$ comes with
an accompanying power of $1/M$. At most one power of $M$ is supplied
by the prefactors, so all powers of ${\bf Q}^2$ higher than the
first are irrelevant in the continuum limit. 

In order to compare our results to the continuum calculation of
Ref.~\cite{thornfreedom}, we must examine the limits $u\to 1$
and $M\to\infty$. The behavior of the first two terms of
Eq.~(\ref{pi2I}) in this limit is transparent once we use the identity
(\ref{uto1-u}).
The continuum limit of the last term requires a bit more analysis.
First, as mentioned in the previous paragraph, we only need
keep two terms in the expansion of the exponential:
\begin{eqnarray}
{g^2N_c\over4\pi^2}\left\{\sum_{l=1}^{M-1}{(2M-l)^2\over l}\sum_{k=1}^\infty
u^k
\left[{u\over Mk+l}-{1\over Mk}\right]
+{{\bf Q}^2\over2T_0}
\sum_{l=1}^{M-1}{(2M-l)^2\over M^3}\sum_{k=1}^\infty
{u^{k+1}(k+1)\over(k+l/M)^2}\right\}.
\end{eqnarray}
The sums over $l$ can be approximated using the 
Euler-Maclaurin summation formula, see Eq.~(\ref{EulerMaclaurin}),
as long as $F$ is not singular at the endpoints of integration.
Clearly terms with $1/l$ in the summand must be treated separately,
which is easily handled using (\ref{psisum}).
Applying these formulae, we find (for large $M$ but arbitrary $u$),
\begin{eqnarray}
&&\hskip-.5in\sum_{l=1}^{M-1}{(2M-l)^2\over l}\sum_{k=1}^\infty
u^k
\left[{u\over Mk+l}-{1\over Mk}\right]\nonumber\\
&=&-\sum_{k=1}^\infty{u^{k+1}\over k}
\sum_{l=1}^{M-1}{(2-l/M)^2\over k+l/M}
+4M(1-u)\ln(1-u)
\left[\psi(M)+\gamma-{7\over8}+{7\over8M}\right]\nonumber\\
&\sim&-\sum_{k=1}^\infty{u^{k+1}\over k}M\left[{7\over3k}+(k+2)^2
\left(\ln\left(1+{1\over k}\right)
-{1\over k}+{1\over2k^2}\right)-{1\over3k}-{2\over k^2}
-{1\over2M}\left({4\over k}+{1\over k+1}\right)\right]\nonumber\\
& &\mbox{}+4M\left[\ln M+\gamma-{7\over8}\right](1-u)\ln(1-u)\nonumber\\
&&\hskip-.5in{{\bf Q}^2\over2T_0}
\sum_{l=1}^{M-1}{(2M-l)^2\over M^3}\sum_{k=1}^\infty
{u^{k+1}(k+1)\over(k+l/M)^2}\nonumber\\
&\sim&{{\bf Q}^2\over2T_0}\sum_{k=1}^\infty
u^{k+1}\left[{7\over3k}-{4\over3k^3}
-2(k+1)(k+2)\left\{\ln\left(1+{1\over k}\right)-{1\over k}+{1\over2k^2}
-{1\over3k^3}\right\}\right]
\end{eqnarray}
The continuum limit also requires $1-u\sim Q^2/2MT_0=({\bf
Q}^2-2mME)/2MT_0$,
so $u$ may be set to unity in all nonsingular
terms without a prefactor of
$M$. Then the above terms simplify to
\begin{eqnarray}
&&\hskip-.5in\sum_{l=1}^{M-1}{(2M-l)^2\over l}\sum_{k=1}^\infty
u^k
\left[{u\over Mk+l}-{1\over Mk}\right]\nonumber\\
&\sim&\left({\alpha\over6}-{\pi^2\over3}\right)M+{\pi^2\over3}+{1\over2}
+{Q^2\over2T_0}\left(\beta+{7\over3}\right)+{2Q^2\over T_0}
\left(\ln M+\gamma-{35\over24}\right)\ln(1-u)\nonumber\\
&&\hskip-.5in{{\bf Q}^2\over2T_0}
\sum_{l=1}^{M-1}{(2M-l)^2\over M^3}\sum_{k=1}^\infty
{u^{k+1}(k+1)\over(k+l/M)^2} \sim
{{\bf Q}^2\over2T_0}\left[-{7\over3}\ln(1-u)+\delta\right]
\end{eqnarray}
where we have defined
\begin{eqnarray}
\alpha&=&12\zeta(3)-6\sum_{k=1}^\infty{(k+2)^2\over k}
\left(\ln\left(1+{1\over k}\right)-{1\over k}+{1\over2k^2}\right) \\
\beta&=&{7\pi^2\over18}+\sum_{k=1}^\infty
{k+1\over k}\left[(k+2)^2
\left(\ln\left(1+{1\over k}\right)-{1\over k}+{1\over2k^2}\right)
-{1\over3k}-{2\over k^2}\right]\\
\delta&=&
-{4\over3}\zeta(3)
-2\sum_{k=1}^\infty (k+1)(k+2)\left[\ln\left(1+{1\over k}\right)-{1\over k}+{1\over2k^2}
-{1\over3k^3}\right]
\end{eqnarray}
which can be numerical evaluated: 
\begin{equation}
\alpha\approx1.188, \qquad \beta\approx1.991, \qquad 
\delta\approx0.633.
\end{equation}

Putting all this together, we get for the continuum limit,
\begin{eqnarray}
\Pi_{2}
&=&M{g^2N_c\over24\pi^2}\left[\alpha+{\pi^2\over6}\right]
+{g^2N_c\over4\pi^2}\left[{\pi^2\over3}
+{1\over2}-{{\bf Q}^2\over T_0}{7\delta\over6}\right]
\nonumber\\
&&\qquad+{g^2N_c\over16\pi^2}{Q^2\over T_0}\left[8\ln M+8\gamma
-{22\over3}\right]\ln{Q^2\over2MT_0}
+{g^2N_c\over24\pi^2}{Q^2\over2T_0}[6\beta+1]
\label{pi2result}
\end{eqnarray}
Remembering that our $\Pi_2$ is a factor of $-Q^2/T_0$ times
that defined in  Ref.~\cite{thornfreedom}, we find agreement
for the coefficient of $\ln Q^2$, provided we identify
$2MT_0$ with $\Lambda^2$.  The first two groups of
terms on the r.h.s.\ of Eq.~(\ref{pi2result}) violate
important symmetries and
must be removed by explicit counter-terms. The term linear in
${\bf Q}^2$ violates Galilei invariance and the momentum
independent terms imply a finite gluon 
mass squared in perturbation theory, thus violating
Poincar\'e invariance. Note that the necessary counter
terms are low order polynomials in both the transverse momentum
and in the discretized $P^+$. Without the tadpole
contributions the discretization used in the text would
have required counter-terms with logarithmic $M$ dependence.
But we found (at least at one-loop) that the tadpoles could be designed to 
eliminate the need for counter-terms. Then the fact that it
preserves Galilei invariance makes it the superior choice.


\end{document}